\documentclass[12pt,preprint]{aastex}

\shorttitle{\feii\ and \hh\ Study of Galactic SNRs}
\shortauthors{Lee et al.}

\usepackage{amsmath}
\usepackage{amsfonts}
\usepackage{amssymb}
\usepackage{color}

\newcommand{\ergscm}{{\rm erg}~{\rm s}^{-1}~{\rm cm}^{-2}}

\newcommand{\msun}{\ensuremath{M_{\sun}}}
\newcommand{\lsun}{\ensuremath{L_{\sun}}}
\newcommand{\kms}{\ensuremath{{\rm km~s}^{-1}}}
\newcommand{\feii}{[\ion{Fe}{2}]}

\newcommand{\pii}{[\ion{P}{2}]}

\newcommand{\hh}{H$_2$}

\newcommand{\hii}{H {\small II}}

\begin{document}

\title{Near-infrared \feii\ and \hh\ Emission-line Study of
Galactic Supernova Remnants in the First Quadrant}

\author{
Yong-Hyun Lee\altaffilmark{1},
Bon-Chul Koo\altaffilmark{1},
Jae-Joon Lee\altaffilmark{2},
Michael G. Burton\altaffilmark{3,4},
and Stuart Ryder\altaffilmark{5,6}
}

\email{yhlee@astro.snu.ac.kr}
\altaffiltext{1}{Department of Physics and Astronomy, Seoul National University,
Seoul 151-742, Republic of Korea}
\altaffiltext{2}{Korea Astronomy and Space Science Institute,
Daejeon 305-348, Republic of Korea}
\altaffiltext{3}{School of Physics, University of New South Wales,
Sydney, NSW 2052, Australia}
\altaffiltext{4}{Armagh Observatory and Planetarium, College Hill, Armagh, BT61 9DG,
Northern Ireland, UK}
\altaffiltext{5}{Australian Astronomical Observatory,
P.O. Box 915, North Ryde, NSW 1670, Australia}
\altaffiltext{6}{Department of Physics and Astronomy, Macquarie University,
Sydney, NSW 2109, Australia}

\begin{abstract}
We report the detection of
near-infrared (NIR) \feii\ (1.644~\micron) and
\hh\ 1--0 S(1) (2.122~\micron) line features
associated with Galactic supernova remnants (SNRs) in the first quadrant
using two narrowband imaging surveys, UWIFE and UWISH2.
Among the total of 79 SNRs fully covered by both surveys,
we found 19 \feii-emitting and 19 \hh-emitting SNRs,
giving a detection rate of 24\% for each.
Eleven SNRs show both emission features.
The detection rate of \feii\ and \hh\ peaks
at the Galactic longitude ($l$) of 40\degr--50\degr\ and 30\degr--40\degr,
respectively,
and gradually decreases toward smaller/larger $l$.
Five out of the eleven SNRs emitting both emission lines
clearly show an ``\feii--\hh\ reversal,''
where \hh\ emission features are found outside the SNR boundary
in \feii\ emission.
Our NIR spectroscopy shows that
the \hh\ emission originates from collisionally excited \hh\ gas.
The brightest SNR in both \feii\ and \hh\ emissions is W49B,
contributing more than 70\% and 50\% of the total
\feii\ 1.644~\micron\ ($2.0 \times 10^4~\lsun$) and
\hh\ 2.122~\micron\ ($1.2 \times 10^3~\lsun$) luminosities 
of the detected SNRs.
The total \feii\ 1.644~\micron\ luminosity of our Galaxy is
a few times smaller than that expected from the SN rate
using the correlation found in nearby starburst galaxies.
We discuss possible explanations for this.
\end{abstract}


\section{Introduction} \label{sec-int}
In our Galaxy, there are about 300 known supernova remnants
\citep[SNRs;][]{gre14}.
They show diverse morphology and properties,
which are either inherited from the supernova (SN) explosion
or acquired through the interaction with their ambient medium.
For massive core-collapse SNRs,
the ambient medium could be the circumstellar medium (CSM)
that their progenitor stars ejected, 
or the interstellar medium (ISM) shaped by the progenitors
through their stellar winds and/or ultraviolet (UV) photons,
or even the molecular clouds (MCs) from which their progenitors had formed.
Therefore, knowing the environment is not only essential
for understanding the morphology and evolution of SNRs,
but is also helpful
for exploring the connection among SNRs, SNe, and progenitor stars.

A useful tool to study the SNR environment and the interaction of SNRs with it
is the near-infrared (NIR; 1--5~\micron) waveband.
Most SNRs are located in the Galactic plane
where the interstellar extinction is large,
so the NIR waveband has a great advantage over the optical waveband.
In the NIR waveband,
there are two prominent lines tracing shocks propagating into dense media:
(1) the 1.644~\micron\ ($a{^4D_{7/2}} \rightarrow a{^4F_{9/2}}$; 
also 1.257~\micron\ from $a{^4D_{7/2}} \rightarrow a{^6D_{9/2}}$)
forbidden line of single ionized Fe (\feii) and
(2) the 2.122~\micron\ ($v$ = 1--0 S(1)) line of molecular hydrogen (\hh).
The \feii\ lines trace  
radiative, fast ($v_{\rm s}=50$--$300~\kms$) $J$-type shocks. 
In radiative $J$-type shocks propagating through either atomic or molecular gas,
an extended region of partially ionized gas at 
constant temperature (6000--8000 K) is developed
in the postshock relaxation layer
\citep[][and references therein]{shu79,hol89,koo16b}.
In this temperature plateau region,
where the gas is photoionized by the strong UV radiation
generated from the hot gas just behind the shock front, 
Fe is mainly in Fe$^{+}$ and the electron density is high.
Hence, the NIR \feii\ lines,
having excitation temperatures of $\sim 12{,}000$ K,
can be easily excited.  
In SNRs, \feii\ 1.644~\micron\ and 1.257~\micron\  
lines are indeed much stronger than H recombination lines 
\citep[e.g.,][]{oli89,mou00,koo15}. 
The \hh\ 2.122~\micron\ line mainly traces slow ($v_{\rm s}<50~\kms$)
nondissociative $C$-type shocks \citep{dra80,che82,dra82}. 
In $C$-type shocks
propagating through molecular gas of low fractional ionization, 
the temperature of the shocked gas is $\lesssim 3000$~K and 
\hh\ is not dissociated,
so that strong \hh\ lines from rovibrational transitions
are emitted in the NIR waveband.
In principle, \hh\ lines can also be emitted from 
dissociative $J$-type shocks if the shock has swept 
up enough column density, i.e., $N_{\rm H} \gtrsim 10^{21-22}$ cm$^{-2}$, 
and \hh\ molecules are re-formed \citep{hol79,hol89,neu89}.
Another important excitation mechanism of \hh\ is
absorption of far-UV photons ($h\nu=11.2$--$13.6$ eV).
The radiative cascade of the excited \hh\ downward to the ground state
also produces strong \hh\ lines \citep[e.g., ][]{bla76,bla87}.
Because the shocked hot plasma is a strong UV and X-ray emitter,
we might also expect \hh\ lines excited by
UV fluorescence around the remnant.

NIR \feii\ and \hh\ emission lines have been studied extensively
in several bright Galactic SNRs: 
Kepler \citep{oli89,ger01},
G11.2$-$0.3 \citep{koo07,moo09},
3C 391 \citep{rea02},
W44 \citep{rea05},
3C 396 \citep{lee09},
W49B \citep{keo07},
Cygnus loop \citep{gra91},
Cassiopeia A \citep{ger01,lee17,koo18},
Crab nebula \citep{gra90,loh11},
IC 443 \citep{tre79,gra87,bur88,bur90a,kok13},
MSH 15-52 \citep{sew83},
RCW 103 \citep{oli90,bur93}.
According to these studies, SNRs bright in \feii\ emission lines
may be divided into two groups \citep{koo14}:
(1) young SNRs interacting with their dense CSMs
(e.g., G11.2$-$0.3, W49B, Cassiopeia A) and
(2) middle-aged SNRs interacting with dense atomic gas or MCs
(e.g., IC 443, W44).
In young core-collapse SNRs, such as Cassiopeia A and G11.2$-$0.2,
strong \feii\ emission from shocked SN ejecta,
with very high expansion velocities ($\gtrsim 1000~\kms$),
has also been detected \citep{ger01,moo09,koo13,lee17}.
The \hh\ emission, on the other hand,
arises from slow, nondissociative $C$-type shocks,
so that it has been observed mainly from
middle-aged SNRs interacting with MCs.
In some SNRs,
e.g., G11.2$-$0.3, 3C 391, W44, 3C 396, W49B, IC 443, and RCW 103,
both \feii\ and \hh\ lines have been detected, 
indicating a complex structure of the ambient medium.

The observed NIR \feii\ and \hh\ emission lines around SNRs
are thought to be mostly produced in SNR shocks,
which could be either fast $J$-type and slow $C$-type
depending on the SNR properties and environment. 
But for some SNRs, their excitation mechanism
as well as their nature are uncertain.
A long-standing problem is to explain the 
extended \hh\ emission features observed 
outside the \feii/radio/X-ray boundary of several Galactic SNRs:
G11.2$-$0.3 \citep{koo07},
3C 396 \citep{lee09},
W49B \citep{keo07},
Cygnus loop \citep{gra91}, and
RCW 103 \citep{oli90,bur93}.
If the shocks are driven by the same SN blast wave, 
we expect the \hh\ filaments to be closer than the \feii\ filaments
to the explosion center,
because in general the former is from slower $C$-type shocks
whereas the latter is from fast $J$-type shocks.
Therefore, this ``\feii--\hh\ reversal'' feature observed in SNRs 
requires an explanation.
Several explanations have been proposed, e.g., 
UV fluorescence excitation, X-ray heating, magnetic precursor,
reflected shock, and complex projection effect
\citep[e.g.,][]{oli90,gra91,bur93,keo07,koo07,lee09},
but the feature remains poorly understood.

Because the two NIR emission lines are closely associated with SNR shocks,
we expect some connection between
the characteristics of these lines and
SN activity in galaxies \citep[e.g.,][]{gre91}.
In particular, because NIR \feii\ emission lines are 
bright in SNRs but faint in \hii\ regions, 
we expect a correlation between the \feii\ luminosity
and the SN rate of a galaxy 
\citep[][and references therein]{gra87,koo15}.
Indeed, some studies have shown a strong correlation 
between \feii\ 1.257/1.644~\micron\ luminosity and SN rates
in bright external galaxies,
\citep[e.g.,][]{mor02,alo03,ros12}.
However, in nearby galaxies where we can resolve individual SNRs,
it is found that about 70--80\% of the \feii\ flux arises from 
diffuse emission of unknown origin and that only 
a small fraction of the SNRs are bright in \feii\ emission \citep{alo03}.
In order to better understand the relation between \feii\ luminosity and SN rate,
therefore, we need to understand the origin of the diffuse \feii\ emission
and the population of \feii-bright SNRs.

In this paper, we carry out a systematic study of SNRs 
using two recent \feii\ and \hh\ narrowband imaging surveys
covering the first quadrant of our Galaxy
\citep{fro11,lee14}.
Using these systematic and unbiased NIR emission surveys,
together with NIR spectroscopic observations,
we investigate the environment and nature of the SNRs.
This paper is organized as follows.
In Section~\ref{sec-obs},
we outline the NIR imaging and spectroscopic observations
and their data reduction/analysis.
The results from the imaging surveys are presented in
Section~\ref{sec-res} together with  
the spectroscopic results for four Galactic SNRs.
We discuss the environment and nature of the SNRs
in Section~\ref{sec-dis},
and the summary of this paper is given in Section~\ref{sec-sum}.

\section{Observations and Data Analysis} \label{sec-obs}
\subsection{Near-infrared Narrowband Imaging Surveys} \label{sec-obs-img}
\subsubsection{Brief description of the surveys} \label{sec-obs-img-des}
The \feii\ and \hh\ narrowband imaging surveys that constitute
UWIFE and UWISH2 were carried out with the Wide-Field Camera (WFCAM)
at the United Kingdom Infrared Telescope
\citep[][]{fro11,lee14}.
The WFCAM consists of four separated Rockwell Hawaii-II
$2048 \times 2048$ detectors with a pixel scale of $0\farcs4$
and each detector spaced 94\% of the imaging area apart \citep{cas07}.
The basic observing unit of a single WFCAM tile is composed of
a $2\times2$ macrostepping sequence in order to fill this spacing,
so that the tile covers a continuous field of view of 0.75 square degrees.
An additional $2\times2$ microstepping sequence with an interlacing technique
has been used during the entire surveys in order to prevent undersampling
at good seeing conditions (less than 0\farcs8).
Therefore, the final stacked images provide $0\farcs2$ pixel sampling.
The single exposure time per frame is 60 s,
but the $2\times2$ microstepping and three jittering observations
result in a total integration time per pixel ($0\farcs4$) of 720 s.
The \feii\ narrowband filter used in the UWIFE survey
has a mean wavelength ($\lambda_{0}$) of $1.645~\micron$ and
a bandwidth ($\Delta \lambda$) of 284 \AA,
while the \hh\ 1--0 S(1) narrowband filter used in the UWISH2 survey
has $\lambda_{0}=2.122~\micron$ and $\Delta \lambda=$211 \AA.
The detailed descriptions for both filters are listed in
Table~\ref{tab-filter}.
Both surveys fully cover the first Galactic quadrant of
$7^{\circ} < l < 62^{\circ}$ and $|b| < 1\fdg3$.
The median seeing of the UWIFE and UWISH2 surveys is
$0\farcs8$ and $0\farcs7$, respectively,
and the typical rms noise level of both surveys goes down to
(1.4--1.6)$\times10^{-17}~{\rm erg}~{\rm s}^{-1}~{\rm cm}^{-2}~{\rm pixel}^{-1}$
in the unbinned image with a $0\farcs2$ pixel scale.
More detailed information is provided in
\citet{lee14} for the UWIFE survey and
\citet{fro11} for the UWISH2 survey.

\subsubsection{Data Reduction and Continuum Subtraction} \label{sec-obs-img-sub}
All WFCAM data are preprocessed by
the Cambridge Astronomical Survey Unit
and are distributed through a dedicated archive
hosted by the Wide Field Astronomy Unit.
During the process,
astrometric and photometric calibration have been made with 
the Two Micron All Sky Survey (2MASS) point-source catalog \citep[][]{skr06}.
A detailed description of the process is presented in \citet{dye06}.

To search for extended \feii/\hh\ emission features more efficiently,
we performed continuum-subtraction
using the broad {\it H}-/{\it K}-band images
obtained as part of the UKIDSS GPS \citep{luc08}.
We first smoothed whichever image was observed in the better seeing
in order to match their point-spread functions (PSFs),
after which the broadband image was scaled to match their fluxes.
In order to suppress the residuals from bright stars as much as possible,
we performed PSF-fitting photometry by using an empirical PSF model
constructed from well-sampled, bright reference stars.
For this, we utilized {\sc StarFinder},
an IDL-based code for deep analysis of stellar fields \citep{dio00}.
Then, bright stars in both narrow- and broadband images
are removed using flux-scaled PSF models.
The final continuum-subtracted \feii/\hh\ images are obtained from
image-to-image subtraction of
the ``bright point source removed'' narrow- and broadband images.
The detailed processing steps of this continuum-subtraction is
described further in Section 3.3 of \citet{lee14}.

\subsubsection{Identification and Flux Measurement} \label{sec-obs-img-ide}
Among the 294 known Galactic SNRs \citep{gre14},
79 SNRs are fully covered by both surveys
and are listed in Table~\ref{tab-detect}.
From the continuum-subtracted images,
we identified \feii\ and \hh\ emission features around the SNRs by eye.
We found various morphologies of the emission lines
that seem to be associated with the SNRs.
We also cross-checked our results with
published catalogs, which were obtained by an automated code
\citep{fro15}.
They reported the detection of \hh\ lines toward 30 Galactic SNRs,
11 of which are either partially covered in the UWIFE survey or
outside of the UWIFE survey area.

In order to measure the fluxes of the identified emission features
using the continuum-subtracted images,
we first performed median-filtering with a ``window'' size of 10 pixels
and masked the star residuals around saturated stars
with $H$-/$K_{\rm s}$-band magnitudes of $> 14$ mag using the 2MASS catalog
to prevent artifacts from dominating the total fluxes
(e.g., hot pixels from cosmic-ray hits, residuals around a saturated star, etc.;
see \citealt{fro15} for details).
We then measured the fluxes 
with an appropriate circular or elliptical aperture
encircling the emission features.
The local background level was estimated from 
an annulus with a radius 1.2--1.5 times larger than the source region.
The total flux ($F$) of \feii\ and \hh\ line emission is derived from
\begin{equation}
F =
F_{0} \left( \dfrac{\rm DN}{t_{\rm exp}} \right) 10^{-0.4 \cdot m_{\rm zp}},
\label{eq-1}
\end{equation}
where $F_{0}$ is the total in-band flux of each narrowband filter for Vega
(see Table~\ref{tab-filter}),
DN is the total digital number of the target during a 60 s exposure time,
and $t_{\rm exp}$ is the effective exposure time fixed to 60 s.
$m_{\rm zp}$ is a zero-point magnitude of the target image
corresponding to the magnitude at
${\rm DN}/t_{\rm exp} = 1~{\rm counts~s}^{-1}$,
which is established by comparing the bright, isolated stars in the field
with the 2MASS $H$-/$K_{\rm s}$-band point-source catalog.
%
The uncertainty of the total flux is derived from the quadrature sum of
(1) the absolute calibration uncertainty and
(2) the standard deviation of the background variation.
Since the background level around the source is quite stable,
most of the flux error arises from the calibration uncertainty.
The typical calibration error,
derived from the uncertainty of the zero-point magnitude over the survey data,
is 0.06 mag for UWIFE and 0.04 mag for UWISH2,
corresponding to 6\% and 4\% of the total fluxes, respectively.
Even in the worst cases, 
the total uncertainty does not exceed 10\% of the total flux.
The {\it H}- and {\it K}-band images used for the continuum-subtraction
(see Section~\ref{sec-obs-img-sub})
include a small portion of \feii\ and \hh\ emission lines,
so the flux measured from the continuum-subtracted image
is roughly 10\% smaller than the true \feii/\hh\ flux,
corresponding to the bandwidth ratios of the narrowband and the broadband filters.
In order to compensate for such ``leakage of \feii/\hh\ flux,''
we multiplied the derived \feii\ and \hh\ fluxes by a factor of 
1.15 (\feii) and 1.10 (\hh), respectively.
These two correction factors were derived from the assumption that
the only emission line in the {\it H} band is the \feii\ 1.644~\micron\ line
and in the {\it K} band, the \hh\ 2.122~\micron\ line.

\subsection{Near-infrared Spectroscopy} \label{sec-obs-spt}
We carried out long-slit NIR spectroscopic observations for four Galactic SNRs
(G11.2$-$0.3, Kes 69, Kes 73, and 3C 391)
in which the bright \feii\ and \hh\ emission lines are detected,
in order to investigate their excitation mechanisms and origins.
These observations were performed with the Infrared Imager and Spectrograph 2
\citep[IRIS2;][]{tin04}
on the 3.9 m Anglo-Australian Telescope,
which provides a spectral resolution of 2200--2500
with a pixel scale of 0$\farcs$45 per pixel
in each of the $J$, $H$, and $K$ bands.
The observations were done in 2012 April 7--8 for G11.2$-$0.3,
and in 2011 June 27 for the other SNRs.
The position angle of the slit is $0\degr$,
i.e., from the north to south direction. 
All observations were performed with multiple on--off nodding sequences
in order to remove night-sky airglow emission lines within the wavebands.
The observation logs are listed in Table~\ref{tab-log}.

We followed the general data reduction procedure.
First, all of the observed raw spectra were subtracted by the dark frame
and then divided by a normalized flat frame.
Using more than a dozen bright OH airglow emission lines,
we derived the two-dimensional wavelength solution
that gives a 1$\sigma$ uncertainty of 2--3\AA\ over the wavelength coverage.
The night-sky airglow emission lines
together with thermal background continuum in the $K$ band
was subtracted by the off-position frame.
Finally, we performed photometric calibration
by comparing the observed spectrum of an A0V type standard star
with the Kurucz model spectrum\footnote{\url{http://kurucz.harvard.edu/}}.
The 1$\sigma$ uncertainty of the absolute flux calibration
reaches up to 30\% of its flux,
due to centering of standard star observations;
however, the relative uncertainty within each band is very robust.

\section{Results} \label{sec-res}
\subsection{[Fe II]/H$_2$ SNRs in the UWIFE and UWISH2 Surveys} \label{sec-res-img}
\subsubsection{Catalog and Some Statistics} \label{sec-res-img-cat}
Among the 79 SNRs covered by these surveys,
we have identified 19 SNRs with \feii\ emission features in the UWIFE survey
and 19 SNRs with \hh\ emission features in the UWISH2 survey.
Eleven of them show both emission features.
The identified SNRs are marked by ``Y'' in the last two columns of 
Table~\ref{tab-detect}, and their \feii\ and/or \hh\ images are 
presented in Figures~\ref{fig-img1-1}--\ref{fig-img3}:
SNRs with both \feii\ and \hh\ lines in Figure~\ref{fig-img1-1}, 
SNRs with \feii\ lines only in Figure~\ref{fig-img2}, and
SNRs with \hh\ lines only in Figure~\ref{fig-img3}. 
Among the 19 \feii-line-detected SNRs,  
six had been known from previous NIR imaging and spectroscopic observations:
G11.2$-$0.3 \citep{koo07}, G21.5$-$0.9 \citep{zaj12},
3C 391 \citep{rea02}, W44 \citep{rea05},
3C 396 \citep{lee09}, and W49B \citep{keo07}. 
The rest (13 out of 19; $\sim 70$\%) are newly discovered in our survey.
Among the 19 \hh-emitting SNRs,
five of them had been known from previous studies:
G11.2$-$0.3 \citep{koo07},
3C 391 \citep{rea02}, W44 \citep{rea05},
3C 396 \citep{lee09}, and W49B \citep{keo07}.
The rest (14 out of 19; $\sim 70$\%) are new discoveries.
The detection rate is 24\% for each survey.  
For comparison, 30--50\%\ of radio SNRs in M82 and NGC 253 are detected in 
\feii\ emission \citep{alo03}.
The lower detection rate in the Milky Way might be partly due to 
the high interstellar extinction in the galactic disk. 
On the other hand, the high detection rate in external galaxies could be 
because the target SNRs are radio-bright ones.
As we will see in the next paragraph,
the radio-bright SNRs have high detection rates in \feii\ emission.

Figure~\ref{fig-hist1} shows the \feii\ and \hh\ detection rates
as a function of Galactic longitude, SNR type, and 1 GHz flux density.
The distribution of the 79 SNRs falling in the survey area is shown 
by the black empty histogram,
while those of the \feii- and \hh-emitting SNRs are 
presented as green and red hatched histograms, respectively.
The detection rate in each bin is also overplotted 
as green diamonds and red triangles, respectively.
In the left panel, we see that 
the distribution of the 79 SNRs peaks at longitude $l=10$\degr--20\degr\
and gradually decreases toward high Galactic longitude.
The small number in the bin at $l=0$\degr--10\degr\ is due to
the limited coverage of the survey (7\degr$ < l < $62\degr).
The detection rate is below 20\% at small $l$ 
and increases toward large $l$, reaching  
a maximum of 50\%\ at 40\degr--50\degr\ for the \feii\ line
and at 30\degr--40\degr\ for the \hh\ line.
It then decreases again to larger $l$.
The low detection rate at small $l$ might be due to
the large interstellar extinction toward this direction,
while the low detection rate at large $l$ could be due to
the relatively diffuse environment there.
%
The middle panel of the figure shows that 
most (62 out of 79) of the SNRs in the survey area are 
shell-type SNRs and that their \feii/\hh\ detection rate is $\sim 20\%$.
The detection rate for the composite SNRs is considerably 
higher than this, i.e., $\sim 50\%$, while for the filled-centered SNRs,
the small number limits statistics.  
Alongside the three radio morphological types,
we also show the detection rate of the mixed-morphology (MM) SNRs.
MM SNRs, which are also known as ``thermal-composite'' SNRs,  
show filled-centered thermal X-ray emission
surrounded by radio shells, and 
most of them show evidence for interaction with MCs 
\citep{rho95,rho98,koo16a}.
There are six SNRs known to be MM types plus 
one MM SNR candidate in our survey area.
(Five of them are shell type and two are composite type in radio.)
The detection rate of MM SNRs is very high, i.e., 
$\sim 90$\% ($=6/7$) in \feii\ and $\sim 60$\% ($=4/7$) in \hh\ emission.
The only MM SNR not detected in the \feii\ line is G33.6$+$0.1 (Kes 79),
which is at a distance of 7.5 kpc \citep{gia09},
so that the nondetection could be due to large extinction.  
The high detection rate of MM SNRs is consistent with the consensus that
these SNRs are in dense environments (see Section~\ref{sec-res-img-lum}). 
%
In the right panel of the figure,
we see that the \feii- and/or \hh-emitting SNRs  
have relatively higher 1 GHz flux density, i.e., 1--300 Jy.
Furthermore, the detection rates are gradually increasing
with the 1 GHz flux density. 
Radio brightness is enhanced 
when SNRs interact with a dense environment, due to higher magnetic fields
and/or higher relativistic electron densities. 
So, this apparent correlation could also be due to their dense environment.

\subsubsection{Morphology and Luminosity} \label{sec-res-img-lum}
In most SNRs, \feii\ emission is confined to thin filaments
or partial shell-like structures 
that are correlated well with bright radio continuum features.
Prototypical SNRs are G28.6$-$0.1, 3C 391, W44, 3C 396, and W49B.  
Such morphology is consistent with the \feii\ emission 
arising from a postshock cooling region
behind a radiative SNR shock propagating into the ambient medium.
Some SNRs, however, show \feii\ emission that does not fit into this category: 
(1) G11.2$-$0.3
where arc-like \feii\ emission features are detected in the central area.
Spectroscopic observations suggest that these \feii\ features are  
associated with fast-moving ($> 1000~\kms$) SN ejecta 
\citep{moo09}.
(2) G21.5$-$0.9 and G41.5$+$0.4,
where the \feii\ emission associated with pulsar wind nebula (PWN) is detected.
In G21.5$-$0.9, the \feii\ emission is confined to thin filaments
surrounding the PWN.
This \feii\ emission had been previously reported by \cite{zaj12}.  
In G41.5$+$0.4,
bright complex \feii\ filaments are detected 
on the central radio structure, which is thought to be a PWN \citep{kap02}.
(3) Kes 73 and G15.9$+$0.2,
where \feii\ emission features are clumpy not filamentary. 
In Kes 73, the \feii\ emission is distributed over the entire remnant
along the radio-bright filaments,
but it is confined to dozens of knotty clumps.
In G15.9$+$0.2, a small ($\sim 4 \arcsec$) \feii\ clump
without enhanced radio continuum emission
is detected near the southwestern boundary of the remnant.

\hh\ emission is also detected mainly toward the bright radio filament/shell
in most SNRs. 
Although the \hh\ emission arises from slow, nondissociative $C$-type shocks
while the radio emission is synchrotron radiation,
the dense environment might cause this correlation.
In some SNRs, however, \hh\ emission has been detected 
beyond the SNR radio boundary.
Prototypical SNRs are G11.2$-$0.3, Kes 73, W44, 3C 396, and W49B
(see Figure~\ref{fig-rev};
a small \hh\ filament is detected outside the western boundary of 
G21.6$-$0.8, too, but their association is not clear).
In G11.2$-$0.3 and W49B, for example, 
we can see extended prominent \hh\ emission well beyond the SNR boundary. 
Note that in the aforementioned five SNRs 
\feii\ emission is also detected
and the offset of \hh\ emission from the \feii\ emission is noticed 
(i.e., \citealt{koo07} for G11.2$-$0.3; \citealt{keo07} for W49B,
\citealt{lee09} for 3C 396, and \citealt{koo14} for W44).
As we mentioned in Section~\ref{sec-int},
this feature, i.e., \hh\ emission farther outside than the \feii\ emission,
is known as the ``\feii--\hh\ reversal'' and needs an explanation.
We will discuss this in Section~\ref{sec-dis-org-h2}.

The observed total \feii\ and \hh\ fluxes of the 27 SNRs are summarized in
Table~\ref{tab-flux}. 
The table also lists the adopted distances and the resulting 
\feii\ 1.644~\micron\ and \hh\ 2.122~\micron\ luminosities. 
The extinction correction has been made by  
using the column densities ($N_{\rm H}$) available in literature,  
assuming the general interstellar dust extinction model \citep{dra03}.
The luminosity ranges are 0.72--$15{,}000~\lsun$ for \feii\ 
and 0.52--$680~\lsun$ for \hh.
W49B is the brightest in either emission. 
The total \feii\ 1.644~\micron\ luminosity is $20{,}000~\lsun$ and
W49B contributes more than 70\% of the total \feii\ 1.644~\micron\ luminosity.
For comparison, the total \hh\ 2.122~\micron\ luminosity is $1200~\lsun$,
half of which is attributed to W49B.
The uncertainty in the measured flux is
less than 10\% (see Section~\ref{sec-obs-img-ide}), but 
the uncertainty of the derived luminosity could be very large,
due to the uncertain distance (and uncertain column density).
For example, the distance to W49B reported in previous studies
varies from 8.0 to 12.5 kpc \citep{loc78,mof94b,zhu14},
and the hydrogen column density is
in the range (4.8--5.3)$\times10^{22}$ cm$^{-2}$ \citep{hwa00,keo07}.
The \feii\ and \hh\ luminosities of W49B, therefore,
are uncertain by a factor of 2.

Even though the two NIR emission lines arise from different types of shocks,
their luminosities seem to be correlated.
This is shown in Figure~\ref{fig-rat}
where the filled symbols represent 11 SNRs
detected in both \feii\ and \hh\ emission
while the empty symbols represent the SNRs detected only in one emission line.
MM SNRs are marked with squares.
The correlation coefficient for the 11 SNRs is $\sim 0.85$.
The brightest SNRs, i.e., W49B, 3C 396, 3C 391, and W44, 
are MM SNRs interacting with a dense ambient medium.
Among the six known MM SNRs in the survey area, 
3C 397 and W51C are exceptions. 
3C 397 is very bright in \feii\ but not detected in \hh, and  
its nature as an MM SNR is uncertain (Section~\ref{sec-dis-org-feii}). 
In W51C, only \feii\ emission is detected. 
\citet{koo97} detected shocked CO but not shocked \hh,
and concluded that the shock is a fast $J$-type shock
and that CO has been reformed but \hh\ has not been yet. 
The nondetection of \hh\ emission also suggests that 
no strong $C$-type shocks are present in this SNR.

\subsection{Spectroscopy of Four SNRs} \label{sec-res-spt}
We carried out NIR spectroscopic observations of four of the SNRs,
i.e., G11.2$-$0.3, Kes 69, Kes 73, and 3C 391,
showing both \feii\ and \hh\ emission features.
Their enlarged \feii/\hh\ images, together with the slit positions,
are shown in Figure~\ref{fig-slit}.
In the $H$- and $K$-band spectra of G11.2$-$0.3 and Kes 69,
we detected \hh\ 1--0 S(1) at 2.122~\micron\ and 
other relatively weak \hh\ lines associated with \hh\ filaments
(Figure~\ref{fig-specks}).
In the $J$- and $H$-band spectra of Kes 73 and 3C 391, on the other hand,
we detected the \feii\ 1.644~\micron\ line and
other weak \feii\ (+ Pa $\beta$) lines associated with \feii\ filaments
(Figure~\ref{fig-specjh}).
We performed a single Gaussian fitting for all of the detected lines
to derive their line-of-sight velocities, line width, and fluxes.
The derived line parameters are listed in
Tables~\ref{tab-h2prop} and \ref{tab-feprop}.
In the following, we describe the results for individual sources.

For G11.2$-$0.3,
we obtained $K$-band spectra along two long slits crossing 
the extended \hh\ filaments.
We obtained 1D spectra at four different positions: N, NE, SE1, and SE2
(Figure~\ref{fig-specks}).
The flux ratio of \hh\ 1--0 S(0) to 1--0 S(1) is $\lesssim 0.2$
in all filaments (Table~\ref{tab-h2prop}),
which is consistent with the collisionally excited \hh\ emission 
at a few 1000 K \citep[$\sim 0.2$;][]{bla87}.
Note that the ratio for UV fluorescence \hh\ ranges from 0.4 to 0.6,
which is higher than that for collisional excitation \citep{bla87}.
Weak/no \hh\ lines from high vibrational levels with $v\geq2$
(2--1 S(3) at 2.07~\micron\ and 2--1 S(1) at 2.25~\micron)
also support the collisional process of the \hh\ filaments.
The $v_{\rm LSR}$ of all \hh\ filaments extended over G11.2$-$0.3
is between $+41~\kms$ and $+47~\kms$ (Table~\ref{tab-h2prop}),
which agrees with the systematic velocity of the remnant
\citep[$+45~\kms$;][]{gre88}.
This implies that
the \hh\ filaments are indeed physically associated with the remnant
despite their large extension.

For Kes 69, we obtained $H$- and $K$-band spectra along a slit
crossing the bright southeastern \hh\ filament. 
We detected bright \hh\ 1--0 S(1) and other weak \hh\ lines
but no ionic lines were detected.
There is some \feii\ emission seen in the image (Figure~\ref{fig-img1-1}),
but the sensitivity of the $H$-band spectrum 
is very weak, and no \feii\ lines were detected.
The ratio of \hh\ lines is again consistent with collisional excitation
(Table~\ref{tab-h2prop}).
The $v_{\rm LSR}$ of the \hh\ filaments is $+57~\kms$
(Table~\ref{tab-h2prop}).
It is known that the SNR is interacting with 
adjacent MCs in this area 
\citep[][and references therein]{zho09}.
\citet{zho09} proposed that
an MC at $+85~\kms$ is associated with the SNR.
Therefore, there is a large difference ($20$--$30~\kms$)
between the velocities of the CO and \hh\ emission. 
We can hypothesize two possible explanations. 
First, the CO emission traces the ambient MC,
whereas the \hh\ emission is from the shocked \hh\ gas.
Hence, the velocity difference may represent
the shock velocity propagating into the MC.
However, the \hh\ filaments are located in the outer boundary of the remnant,
so the shock velocity along the line of sight might be almost negligible.
Second, the systematic velocity of the SNR 
could be $+57~\kms$, not $+85~\kms$.
It is not easy to find an MC associated with an SNR in the inner Galaxy,
due to the confusion by foreground and background CO emission. 
Indeed, we see some CO emission at $\sim 60~\kms$
in the channel maps of \cite{zho09}.
If the systemic velocity of the SNR is $+57\pm3~\kms$
(Table~\ref{tab-h2prop}),
then assuming the flat Galactic rotation model
with the IAU standard rotation constants
($R_{\sun}=8.5$ kpc, $V_{\sun}=220~\kms$),
the kinematic distance to Kes 69 would be $d=4.0\pm0.1$ kpc.
This is considerably smaller than the previous estimation,
e.g., 5.2--5.6 kpc \citep{tia08,zho09,ran18a}.

In Kes 73, one slit is centered on the brightest \feii\ clump
(hereafter ``Knot A'') in the central area.
We have detected a dozen bright \feii\ lines and a weak Pa$\beta$ line
in its $J$- and $H$-band spectra (Figure~\ref{fig-specjh}).
The detection of the H recombination line,
together with the nondetection of the \pii\ 1.189~\micron\ line,
indicates that the knot is either shocked ambient medium or
H-rich SN ejecta with cosmic P/Fe abundance \citep{koo13,lee17}.
The observed central velocity of the line, however, is only $-30~\kms$.
According to SN explosion models,
the H-rich SN ejecta from the progenitor's envelope
show an expansion velocity of more than a few $1000~\kms$
and no less than $300$--$1000~\kms$,
even in significant mixing among the nucleosynthetic layers
\citep[][]{kif06,ham10,won15}.
Therefore, Knot A is probably not shocked SN ejecta
but shocked ambient medium.
Compared to the systematic velocity of the remnant
($+89$ to $+110~\kms$) for Kes 73 \citep{tia08,kil16},
however, it is much ($\sim 100~\kms$) larger,
so that the shocked ambient medium might be the CSM not the ISM.
This is consistent with the clumpy,
rather than filamentary, morphology of the \feii\ emission.
We consider that the observed \feii\ knots are 
dense clumps in the circumstellar wind swept up by the SNR shock.
From the ratios of the \feii\ lines \citep{koo16b},
we found that the electron density of Knot A is $\sim7000$ cm$^{-3}$.

In 3C 391, spectra along two $H$-band slits crossing
the extended \feii\ filaments (Figure~\ref{fig-slit}) were obtained.
We detected the \feii\ 1.644~\micron\ line and, 
in the brightest \feii\ filament in Slit 2 (hereafter ``Spot A''),
we also detected additional \feii\ lines (Figure~\ref{fig-specjh}).
Figure~\ref{fig-3c391} shows the position-velocity diagrams of
the \feii\ 1.644~\micron\ line along the two slits.
The central velocity of the \feii\ filament varies with position
from $-200~\kms$ to $+100~\kms$.
The systemic velocity of 3C 391 is $+100~\kms$ \citep{rea99,kil16,ran17},
so these filaments are blueshifted by $0$--$200~\kms$.
From the ratios of the \feii\ lines \citep{koo16b},
we found that the electron density of Spot A is $\sim3000$ cm$^{-3}$.

\section{Discussion} \label{sec-dis}
\subsection{Nature of the SNRs with \feii/\hh\ Emission} \label{sec-dis-org}
We have detected 19 SNRs with \feii\ and 19 SNRs with \hh\ emission.
Eleven of them are detected in both \feii\ and \hh\ emission.
In this section, we discuss the nature of these SNRs.

\subsubsection{SNRs with \feii\ Emission} \label{sec-dis-org-feii}
According to previous NIR studies of SNRs,
\feii\ emission mainly arises from dense CSM/ISM
swept up by an SN shock \citep[e.g.,][]{gra87,bur93,keo07,koo07,lee09}.
The detection of strong \feii\ emission
in fast-moving ($> 1000~\kms$) SN ejecta were reported
for only two young SNRs, Cassiopeia A and G11.2$-$0.3
\citep{ger01,moo09,koo13,lee17}.
In our survey, we detected \feii\ emission associated with PWN
in two SNRs, i.e., G21.5$-$0.9 and G41.5$+$0.4.
Considering that the PWN is expanding into the SN ejecta,
the emission is likely to be from the shocked SN ejecta.

Among the rest, the \feii\ emission in Kes 73 is likely to be from 
shocked dense clumps in the circumstellar wind (Section~\ref{sec-res-spt}).
Kes 73 is a shell-type SNR with a radius of 2\farcm5,
hosting the anomalous X-ray pulsar 1E 1841$-$045 \citep{hel94,vas97}.
It is believed to be of one of the youngest ($\lesssim 2000$ yr)
Galactic SNRs \citep{kum14,bor17}.
Previous X-ray studies reported that the hydrogen number density of
the surrounding medium ($n_{0}$) is 2 cm$^{-3}$
and that the forward shock velocity ($v_{\rm fs}$) is 1400~\kms\
\citep{kum14,bor17}.
Then, assuming that the velocity of the radiative shock front
propagating into the dense \feii-emitting clumps ($v_{\rm c}$) is 
100--200~\kms, the preshock density of the clump would be
$n_{\rm c} \approx n_{0} (v_{\rm fs}/v_{\rm c})^{2}$
$\sim 100$--$400~{\rm cm}^{-3}$.
Similar dense \feii\ clumps are detected in
the young ($\sim 340$~yr) SNR Cassiopeia A,
where the shock speed is $\sim 5000~\kms$, while 
the velocity of the clumps is $\lesssim 300~\kms$ \citep{che03}.
These clumps are N- and He-rich, 
and are believed to be dense clumps embedded in the smooth wind
ejected during the red supergiant phase of the progenitor
\citep{ger01,lee17}.
We, therefore, suggest that
the \feii\ clumps in Kes 73 are shocked, dense circumstellar clumps
similar to Cassiopeia A and that the Kes 73 SN might be
an SN IIP or IIb/L exploding in the red supergiant stage.
This conclusion is consistent with that of a recent X-ray study
\citep{bor17}.

The four brightest SNRs with $L{\rm ([Fe~II])} \gtrsim 1000~\lsun$
(3C 391, 3C 396, 3C 397, and W49B) are MM SNRs. 
3C 391 and 3C 396 are middle-aged SNRs with ample evidence 
for interaction with MCs 
\citep[e.g.,][]{wil98,che04,lee09,su11}.
In high-resolution radio images,
3C 391 shows a partial shell of 5\arcmin\ radius,
with relatively faint emission extending through the broken shell
in the southeastern part \citep{rey93,mof94a}.
This ``breakout'' morphology,
together with the CO cloud blocking the northeastern area, implies that
SN explosion took place at the edge of an MC \citep{wil98}. 
The detection of two 1720 MHz OH maser spots 
indicates that the remnant is currently interacting with
the surrounding MCs \citep{fra96}.
3C 396, on the other hand,
shows a bright western incomplete shell in radio \citep[e.g.,][]{bec87}
and possesses a central X-ray PWN \citep{har99}.
Previous radio and infrared observations reported
the detection of molecular gas emitting bright \hh\ and CO emission lines
along the western boundary, and found that
they are indeed physically in contact with the SNR \citep{rea06,lee09,su11}.
The SNR is believed to be the remnant of a core-collapse SN
with a $13$--$15~\msun$ B1--B2 progenitor,
with its blast wave currently running into an MC \citep{su11}.

In contrast to these two SNRs, the nature of 3C 397 is uncertain
\citep[e.g.,][and references therein]{koo16a}. 
It has been suggested that
the SNR with a core-collapse SN origin is currently interacting with  
its mother MCs in the western edge of the remnant \citep[][]{saf00,saf05,jia10}.
However, high Ni and Mg abundances from X-ray observations,
together with the lack of a compact source inside the remnant,
suggest that 3C 397 is the result of an SN Ia explosion
\citep[e.g.,][]{che99a,yan13,yam14}.
\cite{koo16a} noted that its IR to X-ray ratio is much smaller than
the other MM SNRs, so they suggested the SN Ia origin.
Although the nondetection of \hh\ emission in our observations
does not support the core-collapse SN origin,
its unusual \feii\ and radio morphology
seems to be shaped by a dense surrounding medium
rather than by an asymmetric SN explosion.
We note that the \feii\ is very bright in the northeastern edge,
but relatively weak in the southwestern edge
where the radio emission is enhanced.
This suggests that the \feii\ emission lines could be 
from Fe-rich SN ejecta.

Finally, the exceptionally high \feii\ luminosity of W49B is puzzling. 
W49B shows a barrel-like morphology in the radio and NIR wavebands,
but has centrally brightened thermal X-ray emission
\citep[e.g.,][]{pye84,mof94b,hwa00,keo07}.
\citet{keo07} argued that 
the barrel-like \feii\ morphology is the result of 
shock interaction with a wind-blown bubble shaped by its WR progenitor,
while \citet{lop13} proposed a jet-driven explosion scenario for the SNR.
The \feii\ emission in W49B could have been enhanced by
its strong X-ray emission.  
W49B is one of the most luminous Galactic SNRs in X-ray and $\gamma$-ray emission
\citep{imm05,abd10}, and this strong radiation field could produce
a partially ionized region emitting \feii\ lines \citep[e.g.,][]{moo88}.
Another possibility is that
the \feii\ emission is from Fe-rich SN ejecta with high Fe abundance.
However, the morphological similarity to the radio continuum,
rather than the X-rays, seems to indicate that
the \feii\ emission is associated with CSM/ISM rather than the SN ejecta.

\subsubsection{SNRs with \hh\ Emission} \label{sec-dis-org-h2}
\hh\ emission is strong evidence for the interaction of the SNR with an MC,
and this ``SNR--MC interaction'' is often thought to be 
indication that the progenitor was a core-collapse SN 
\citep[e.g.,][]{hua86,che99b}.
Massive stars with an initial mass of $\gtrsim 8~\msun$ are born in giant MCs
and end their lives as core-collapse SNe
after $\lesssim 3 \times 10^{7}$ yr.
Unless photoionizing photons and/or stellar winds from the progenitors
have perfectly cleared out the surrounding MCs,
the SNRs will be interacting with the dense MC material.
Early B stars with initial mass 8--12~$\msun$ 
may explode within their parental MCs \citep{che99b}.  
Some well-known SNRs interacting with MCs in our survey area are
Kes 69, 3C 391, and W44 \citep[e.g.,][]{woo77,gre97,wil98}.
In addition to the \hh\ emission,
these SNRs also show OH masers and/or broad CO lines
supporting the presence of SNR--MC interactions \citep{jia10}.
For the remaining SNRs, the detection of \hh\ emission is the first strong 
evidence for their interaction with MCs.
A caveat, however, is that
the \hh\ emission features might not be associated with the SNR,
and a detailed study of each SNR is necessary to confirm the association. 

Among the 19 \hh-emitting SNRs,
11 SNRs show both \feii\ and \hh\ emission lines
(Section~\ref{sec-res-img}).
We can consider an SNR interacting with a clumpy MC,
where the interclump medium is a low-density atomic medium \citep{che99b}.
In such a case, the shock propagating into the dense clumps
could be a nondissociative molecular shock,
while the shock propagating into the interclump medium could be
a radiative atomic shock.
Hence, we see \hh\ emission from shocked clumps
and \feii\ emission from shocked interclump atomic gas.
\citet{che99b} showed that such an interpretation can explain most of the 
observed properties in W44.
We consider that a similar explanation,
i.e., an SNR in an environment where dense molecular gas coexists with atomic gas,
might be applicable to most of the SNRs
showing both \hh\ and \feii\ emission lines.

An interesting phenomenon in these SNRs, however,
is the ``\feii--\hh\ reversal,''
where \hh\ emission features are located outside the \feii\ filaments.
As explained in Section~\ref{sec-int},
we expect \hh\ filaments due to slow $C$-type shocks to be closer to
the explosion center than the \feii\ filaments produced by fast $J$-type shocks.
The \hh\ emission can also originate from the $J$-type shock,
if \hh\ molecules re-form in further downstream
from the \feii-emitting region after the shock passage
\citep{hol79,hol89,neu89}.
Even in this case,
the \hh\ filaments are expected to be inside the \feii\ filaments.
Among the 11 SNRs, five show the ``\feii--\hh\ reversal'' phenomenon:
G11.2$-$0.3, Kes 73, W44, 3C 396, and W49B (Figure~\ref{fig-rev}).
\hh\ filaments with a small offset from \feii\ filaments could be due to 
projection effects or a magnetic precursor,
e.g., Kes 73 and W44.
But in G11.2$-$0.3, for example, the extended \hh\ filaments
are detected far beyond the \feii\ and radio boundary, 
at almost twice the remnant's radius from the geometrical center.
Such \hh\ filaments seem difficult to explain by shock excitation
considering their large distance from the radio or X-ray SNR boundary. 
Another possibility might be the excitation by high-energy photons.
In Section~\ref{sec-res-spt}, however,
we showed that \hh\ flux ratios are consistent with
collisional excitation and not with UV/X-ray excitation.
It is still possible that
we can observe the \hh\ line ratio closer to the collisional excitation case
if the density of molecular gas heated by UV/X-ray radiation is very high 
\citep[$\gtrsim 10^{6}$ cm$^{-3}$;][]{ste89,bur90b}.
In such a case, we expect a line width of $\lesssim 5~\kms$,
corresponding to the typical turbulent velocity of ISM/MCs
\citep{hol89,bur92},
which is much narrower than what we expect for the shock excitation
(i.e., a few $10~\kms$).
Our spectroscopic observation had insufficient spectral resolution
to address this issue,
and high-resolution NIR spectroscopic observations will be needed.

\subsection{\feii\ Luminosity and Supernova Rate} \label{sec-dis-rat}
Since the NIR \feii\ emission is bright in SNRs
but relatively faint in \hii\ regions
\citep[][and references therein]{gra87,koo15},
it has been regarded as a tracer of SN activity in galaxies
\citep{gre91,alo03,ros12}.
In Figure~\ref{fig-feii},
we first compare the \feii\ 1.644~\micron\ luminosity distribution of SNRs
in external galaxies with our results.
Two out of seven external galaxies, LMC and M33,
are normal galaxies in the Local Group,
whereas the rest are nearby starburst galaxies.
It is clear that faint SNRs are missed in external galaxies.
The faintest SNR in nearby galaxies is $\gtrsim 10~\lsun$ (e.g., LMC, M33),
while it is more than an order of magnitude fainter in the Milky Way.
This might be due to the limited sensitivity of extragalactic \feii\ studies.
The contribution of these faint SNRs to the total 
\feii\ 1.644~\micron\ luminosity of a galaxy, however, should be almost negligible.
Figure~\ref{fig-feii} also shows that the brightest SNR in most galaxies 
is not as bright as the SNRs in the Milky Way. 
In the LMC and M33, for example, the \feii\ 1.644~\micron\ luminosity of the 
brightest SNR is $\sim 700~\lsun$,
which is less than that ($960~\lsun$) of the fourth brightest SNR (3C 391)
in the Milky Way. 
In NGC 6946, the \feii\ 1.644~\micron\ luminosities of the two brightest SNRs are
$1800~\lsun$ and $3300~\lsun$ \citep{bru14},
which are comparable to those of the second and third brightest SNRs
(i.e., 3C 396 and 3C 397)
but much fainter than the brightest SNR (W49B) in the Milky Way. 
It is only the two starburst galaxies, M82 and NGC 253,
where we see SNRs as bright as W49B. 
\citet{mor02} already noted that the brightest SNR in M82
is two orders of magnitude brighter than that of M33,
and attributed this large discrepancy to
the different ISM densities (and the different metallicities)
prevailing in different types of galaxy.
In contrast to the Milky Way, however,
the brightest SNR in M82 and NGC 253 only accounts for
3--4\% of the total \feii\ 1.644~\micron\ luminosity associated with the SNRs
\citep{alo03}.

Efforts have been made to find a correlation between
the total \feii\ 1.257/1.644~\micron\ luminosity of galaxies and their SN rates
\citep{mor02,alo03,ros12}.
\citet{mor02} performed \feii\ narrowband imaging surveys toward 
42 optically selected SNRs in M33
and detected only seven \feii-emitting SNRs.
They suggested that this low detection rate (17\%) could be due to either   
the finite duration of \feii-line emitting phase ($\sim 10^{4}$ yr) or
an SNR sample biased in favor of objects evolving in a warm, tenuous ISM.
They also showed that
the \feii\ 1.644~\micron\ luminosity is strongly correlated with
the electron density of the postshock gas and also
the metallicity of the shock-heated gas.
On the basis of these results, they provided an empirical relation
allowing the determination of the current SN rate of starburst galaxies
from their total \feii\ 1.644~\micron\ luminosity.
On the other hand,
\citet{alo03} obtained an HST image of M82 and NGC 253,
and detected \feii\ emission in 30--50\% of radio SNRs
that are thought to be middle-aged SNRs. 
They found that 70--80\% of the total \feii\ 1.644~\micron\ luminosity
arises from diffuse sources without corresponding SNRs
and attributed this diffuse \feii\ emission to unresolved or merged SNRs.
By comparing the total \feii\ 1.644~\micron\ luminosity to the SN rate
derived from the number counts of radio SNRs,
they derived a linear relationship between these quantities.
More recently, \citet{ros12} investigated the correlation
between the \feii\ 1.257~\micron\ luminosity and
the SN rate in 11 nearby starburst galaxies.
By applying a starburst model
to Br-$\gamma$ equivalent width to individual pixels,
they found a tight correlation between the SN rate
($\nu_{\rm SN}$ in units of $N_{\rm SNe}~{\rm year}^{-1}$) and
the \feii\ 1.257~\micron\ luminosity
($L_{\rm [Fe~II]1.257}$ in units of $\lsun$),
which can be converted to
the \feii\ 1.644~\micron\ luminosity ($L_{\rm [Fe~II]1.644}$)
assuming $F({\rm [Fe~II]~1.257})/F({\rm [Fe~II]~1.644}) = 1.36$
\citep{nus88,deb10}.
Equation~(2) in \citet{ros12}, therefore, can be rewritten as
$$\log (\nu_{\rm SN}) = 0.89 \pm 0.2 \times \log (L_{\rm [Fe~II]1.644}) -36.09 \pm 0.9.$$
This relation appears to be applicable to starburst galaxies with
a total \feii\ 1.644~\micron\ luminosity
between $3\times10^4~\lsun$ and $4\times 10^7~\lsun$. 
If we naively substitute the SN rate of the Milky Way, 
i.e., two to five SNe per century
\citep[][and references therein]{die06,li11,ada13}, to the above equation, 
we obtain an \feii\ 1.644~\micron\ luminosity of (1--3)$\times 10^{5}~\lsun$.
For comparison,
the total \feii\ 1.644~\micron\ luminosity of SNRs from our survey is
$2 \times 10^{4}~\lsun$ (Table~\ref{tab-flux}).
Since our survey covers only 27\% of known SNRs (79 out of 294),
we can simply multiply a factor of 4
to the observed \feii\ 1.644~\micron\ luminosity,
which yields $L_{\rm [Fe~II]1.644} \sim 8 \times 10^{4}~\lsun$.
This is a few times smaller than
the expected total \feii\ 1.644~\micron\ luminosity
inferred from the Galactic SN rate.
It is possible that we have missed \feii\ bright SNRs
either because of our limited coverage in galactic longitude
or because of the large extinction in the Galactic plane. 
On the other hand,
considering that about 70--80\% of the \feii\ 1.644~\micron\ luminosity 
in these starburst galaxies (e.g., M82 and NGC 253) arises from 
diffuse sources without SNR counterparts \citep{alo03},
there could be some significant contribution from
diffuse \feii\ emission in the Milky Way, too.
Or, the equation in \citet{ros12} may not be applicable to
normal galaxies like the Milky Way.
A systematic study of nearby galaxies is needed to explore
the possible relation between the \feii\ luminosity and the SN rate
in normal galaxies.

\section{Summary} \label{sec-sum}
%
We have searched for \feii\ 1.644~\micron\ and
\hh\ 2.122~\micron\ emission-line features
around 79 Galactic SNRs using the UWIFE and UWISH2 surveys.
Bright emission lines with various morphologies were detected
around 27 SNRs.
We also performed NIR spectroscopic observations of four Galactic SNRs
(G11.2$-$0.3, Kes 69, Kes 73, and 3C 391)
showing both \feii\ and \hh\ lines in the surveys,
in order to investigate their excitation mechanisms as well as their origins.
Our main results are listed in the following.

\noindent
1. Among the 79 Galactic SNRs fully covered by the surveys,
we found 19 \feii-emitting and 19 \hh-emitting SNRs
corresponding to a 24\% detection rate for each,
and 11 of them are emitting both \feii\ and \hh\ lines.
Furthermore, more than half of our detections are new discoveries
that have never been reported in previous studies.

\noindent
2. The detection rate reaches up to $\sim 50$\%
at $l=40$\degr--50\degr\ for \feii\ and at $l=30$\degr--40\degr\ for \hh,
and gradually decreases toward lower/higher $l$.
The low detection rate at small $l$ might be due to
large interstellar extinction to this direction,
while the low detection rate at large $l$ could be due to
the relatively diffuse environment there.
We also found that the detection rate is very high ($\sim 90\%$) for MM SNRs,
with higher detection rates for SNRs with larger 1 GHz flux densities.
This is consistent with the consensus that
those SNRs are currently interacting with their dense environments,
and that the detection of \feii/\hh\ is another indicator of the SNRs
interacting with their dense surrounding medium.

\noindent
3. The small radial velocities of \feii\ emission features
(with cosmic abundance) detected in both 3C 391 and Kes 73
imply that they are shocked CSM/ISM,
rather than the high-speed, metal-enriched SN ejecta.
The \feii\ morphologies of these two SNRs, however, are very different,
(i.e., diffuse/filamentary \feii\ in 3C 391 vs. small clumpy \feii\ in Kes 73),
and this may be due to different density distributions of
their surrounding medium.
We suggest that the \feii\ clumps in Kes 73 could be  
shocked, dense circumstellar clumps ejected during its red supergiant phase.

\noindent
4. Five bright SNRs (G11.2$-$0.3, Kes 73, W44, 3C 396, and W49B)
emitting both \feii\ and \hh\ lines clearly show an ``\feii--\hh\ reversal;''
\hh\ emission extends outside of the radio and \feii\ emission-line boundary.
In G11.2$-$0.3, the extended \hh\ filaments are detected
at almost twice the remnant's radius from the geometrical center.
Our NIR spectroscopy showed that
they are probably associated with the remnant 
and arise from the collisionally excited \hh\ gas.
The exciting source, however, remains to be explored.

\noindent
5. The total \feii\ 1.644~\micron\ luminosity in our survey is
$2\times10^{4}~\lsun$,
and W49B is responsible for more than 70\% of this.
The total \feii\ 1.644~\micron\ luminosity of our Galaxy,
extrapolated from our observations,
is a few times smaller than that expected from the correlation
between the SN rate of nearby starburst galaxies and
their total \feii\ 1.644~\micron\ luminosities
($8\times10^{4}~\lsun$ vs. (1--3)$\times10^{5}~\lsun$).
This discrepancy could be due to either
the limited coverage of our surveys,
the large extinction in the galactic plane, or
the different interstellar environments
in starburst galaxies and normal galaxies like the Milky Way.

\acknowledgments
This work was supported by Basic Science Research Program
through the National Research Foundation of Korea (NRF)
funded by the Ministry of Science, ICT and Future Planning (2017R1A2A2A05001337)
to B.-C. K.
The UWIFE survey was supported by the Korean GMT Project
operated by the Korea Astronomy and Space Science Institute (KASI).

{}

\clearpage
\begin{deluxetable}{lccccc}
\tabletypesize{\scriptsize}
\tablewidth{0pt}
\tablecolumns{5}
\tablecaption{Specifications of the \feii\ and \hh\ Narrowband Filters
\label{tab-filter}}
\tablehead{
\colhead{Filter}   &
\colhead{BW\tablenotemark{a}}   &
\colhead{$\lambda_{0}$\tablenotemark{b}}   &
\colhead{$\lambda_{\rm iso}$}   &
\colhead{$F_\lambda(\lambda_{\rm iso})$\tablenotemark{c}}	&
\colhead{In-band\tablenotemark{d}}  \\
\colhead{}   &
\colhead{(\AA)}   &
\colhead{($\micron$)}   &
\colhead{($\micron$)}   &
\colhead{(W m$^{-2}$ $\micron^{-1}$)}   &
\colhead{(W m$^{-2}$)}
}
\startdata
[Fe II]	&	284	&	1.645	&	1.666	&	1.15E-9	&	3.27E-11	\\
H$_{2}$	&	211	&	2.122	&	2.122	&	4.66E-10&	9.84E-12	\\
\enddata
\tablenotetext{a}{
Equivalent band width defined by
${\rm BW}=\int R(\lambda)~d\lambda~/~R_{\rm max}$,
where $R_{\rm max}$ is the maximum throughput of
the filter response function ($R(\lambda)$).
}
\tablenotetext{b}{
Mean wavelength of the filter defined by
$\lambda_{0}=\int \lambda R(\lambda) d\lambda / \int R(\lambda) d\lambda$.
}
\tablenotetext{c}{
Zero-point level for the Vega continuum.
It is described by
$F_\lambda(\lambda_{\rm iso})=
\int \lambda F_{\lambda}(\lambda) R(\lambda)~d\lambda~/~
\int \lambda R(\lambda)~d\lambda$,
where $F_{\lambda}(\lambda)$ is
the spectral energy distribution of Vega \citep{rie08}.
$\lambda_{\rm iso}$ denotes the ``isophotal wavelength''
at which the $F_\lambda(\lambda_{\rm iso})$
equals the flux density of the Vega continuum
(see \citet{tok05} and \citet{rie08} for more information).
}
\tablenotetext{d}{
Total in-band flux of the Vega spectrum falling in the passband,
derived from the $F_\lambda(\lambda_{\rm iso})$
multiplied by the equivalent bandwidth (BW).
}
\end{deluxetable}

\clearpage
\begin{deluxetable}{rcclccc}
\tabletypesize{\scriptsize}
\tablewidth{0pt}
\tablecolumns{7}
\tablecaption{\feii\ and \hh\ Detection of the Galactic SNRs \label{tab-detect}}
\tablehead{
\colhead{G-Name}   & \colhead{Other Name}   & \colhead{Size\tablenotemark{a}}   & \colhead{Type\tablenotemark{b}}	&
\colhead{$F_{\rm 1~GHz}$\tablenotemark{c}}	&	\multicolumn{2}{c}{Detection?\tablenotemark{d}}	\\
\cline{6-7}
\colhead{}   &\colhead{}   & \colhead{(arcmin)}   &\colhead{(MM?)}   &
\colhead{(Jy)}	&	\colhead{[Fe II]}  & \colhead{H$_{2}$}
}
\startdata
G$ 7.2+0.2$	&	...			&	12				&	S		&	2.8		&	...	&	...	\\
G$ 8.3-0.0$	&	...			&	5 $\times$ 4		&	S		&	1.2		&	...	&	...	\\
G$ 8.7-0.1$	&	(W30)		&	45				&	S?		&	80		&	Y	&	...	\\
G$ 8.9+0.4$	&	...			&	24				&	S		&	9		&	...	&	...	\\
G$ 9.7-0.0$	&	...			&	15 $\times$ 11	&	S		&	3.7		&	...	&	...	\\
G$ 9.8+0.6$	&	...			&	12				&	S		&	3.9		&	...	&	...	\\
G$ 9.9-0.8$	&	...			&	12				&	S		&	6.7		&	...	&	Y	\\
G$10.5-0.0$	&	...			&	6				&	S		&	0.9		&	...	&	...	\\
G$11.0-0.0$	&	...			&	11 $\times$ 9	&	S		&	1.3		&	...	&	...	\\
G$11.1-1.0$	&	...			&	18 $\times$ 12	&	S		&	5.8		&	...	&	...	\\
G$11.1-0.7$	&	...			&	11 $\times$ 7	&	S		&	1.0		&	...	&	...	\\
G$11.1+0.1$	&	...			&	12 $\times$ 10	&	S		&	2.3		&	...	&	...	\\
G$11.2-0.3$	&	...			&	4				&	C		&	22		&	Y	&	Y	\\
G$11.4-0.1$	&	...			&	8				&	S?		&	6		&	...	&	...	\\
G$11.8-0.2$	&	...			&	4				&	S		&	0.7		&	...	&	...	\\
G$12.0-0.1$	&	...			&	7?				&	?		&	3.5		&	...	&	...	\\
G$12.2+0.3$	&	...			&	6 $\times$ 5		&	S		&	0.8		&	...	&	...	\\
G$12.5+0.2$	&	...			&	6 $\times$ 5		&	C?		&	0.6		&	...	&	...	\\
G$12.7-0.0$	&	...			&	6				&	S		&	0.8		&	...	&	...	\\
G$12.8-0.0$	&	...			&	3				&	C?		&	0.8		&	...	&	...	\\
G$13.5+0.2$	&	...			&	5 $\times$ 4		&	S		&	3.5?		&	...	&	Y	\\
G$14.1-0.1$	&	...			&	6 $\times$ 5		&	S		&	0.5		&	...	&	...	\\
G$14.3+0.1$	&	...			&	5 $\times$ 4		&	S		&	0.6		&	...	&	...	\\
G$15.4+0.1$	&	...			&	15 $\times$ 14	&	S		&	5.6		&	...	&	...	\\
G$15.9+0.2$	&	...			&	7 $\times$ 5		&	S?		&	5.0		&	Y	&	...	\\
G$16.0-0.5$	&	...			&	15 $\times$ 10	&	S		&	2.7		&	...	&	Y	\\
G$16.4-0.5$	&	...			&	13				&	S		&	4.6		&	...	&	...	\\
G$16.7+0.1$	&	...			&	4				&	C		&	3.0		&	...	&	...	\\
G$17.0-0.0$	&	...			&	5				&	S		&	0.5		&	...	&	...	\\
G$17.4-0.1$	&	...			&	6				&	S		&	0.4		&	...	&	...	\\
G$18.1-0.1$	&	...			&	8				&	S		&	4.6		&	Y	&	Y	\\
G$18.6-0.2$	&	...			&	6				&	S		&	1.4		&	...	&	...	\\
G$18.8+0.3$	&	Kes 67		&	17 $\times$ 11	&	S		&	33		&	...	&	...	\\
G$18.9-1.1$	&	...			&	33				&	C?		&	37		&	Y	&	Y	\\
G$19.1+0.2$	&	...			&	27				&	S		&	10		&	...	&	...	\\
G$20.0-0.2$	&	...			&	10				&	F		&	10		&	...	&	...	\\
G$20.4+0.1$	&	...			&	8				&	S?		&	9?		&	...	&	...	\\
G$21.0-0.4$	&	...			&	9 $\times$ 7		&	S		&	1.1		&	...	&	...	\\
G$21.5-0.9$	&	...			&	5				&	C		&	7		&	Y	&	...	\\
G$21.5-0.1$	&	...			&	5				&	S		&	0.4		&	...	&	...	\\
G$21.6-0.8$	&	...			&	13				&	S		&	1.4		&	...	&	Y	\\
G$21.8-0.6$	&	Kes 69		&	20				&	S		&	65		&	Y	&	Y	\\
G$22.7-0.2$	&	...			&	26				&	S?		&	33		&	...	&	...	\\
G$23.3-0.3$	&	W41			&	27				&	S		&	70		&	Y	&	...	\\
G$23.6+0.3$	&	...			&	10?				&	?		&	8?		&	...	&	...	\\
G$24.7-0.6$	&	...			&	15?				&	S?		&	8		&	...	&	...	\\
G$24.7+0.6$	&	...			&	30 $\times$ 15	&	C?		&	20?		&	...	&	Y	\\
G$27.4+0.0$	&	Kes 73		&	4				&	S		&	6		&	Y	&	Y	\\
G$27.8+0.6$	&	...			&	50 $\times$ 30	&	F		&	30		&	Y	&	Y	\\
G$28.6-0.1$	&	...			&	13 $\times$ 9	&	S		&	3?		&	Y	&	...	\\
G$29.6+0.1$	&	...			&	5				&	S		&	1.5?		&	...	&	...	\\
G$29.7-0.3$	&	Kes 75		&	3				&	C		&	10		&	...	&	...	\\
G$30.7+1.0$	&	...			&	24 $\times$ 18	&	S?		&	6		&	...	&	...	\\
G$31.5-0.6$	&	...			&	18?				&	S?		&	2?		&	...	&	...	\\
G$31.9+0.0$	&	3C 391		&	7 $\times$ 5		&	S (Y)	&	25		&	Y	&	Y	\\
G$32.1-0.9$	&	...			&	40?				&	C?		&	?		&	...	&	Y	\\
G$32.4+0.1$	&	...			&	6				&	S		&	0.25?	&	...	&	...	\\
G$32.8-0.1$	&	Kes 78		&	17				&	S?		&	11?		&	Y	&	Y	\\
G$33.2-0.6$	&	...			&	18				&	S		&	3.5		&	...	&	Y	\\
G$33.6+0.1$	&	Kes 79		&	10				&	S (Y)	&	20		&	...	&	...	\\
G$34.7-0.4$	&	W44			&	35 $\times$ 27	&	C (Y)	&	250		&	Y	&	Y	\\
G$35.6-0.4$	&	...			&	15 $\times$ 11	&	S?		&	9		&	...	&	...	\\
G$36.6-0.7$	&	...			&	25?				&	S?		&	1.0		&	...	&	...	\\
G$39.2-0.3$	&	3C 396		&	8 $\times$ 6		&	C (Y)	&	18		&	Y	&	Y	\\
G$40.5-0.5$	&	...			&	22				&	S		&	11		&	...	&	...	\\
G$41.1-0.3$	&	3C 397		&	4.5 $\times$ 2.5	&	S (Y)	&	25		&	Y	&	...	\\
G$41.5+0.4$	&	...			&	10				&	S?		&	1?		&	Y	&	...	\\
G$42.0-0.1$	&	...			&	8				&	S?		&	0.5?		&	...	&	...	\\
G$42.8+0.6$	&	...			&	24				&	S		&	3?		&	...	&	...	\\
G$43.3-0.2$	&	W49B			&	4 $\times$ 3		&	S (Y)	&	38		&	Y	&	Y	\\
G$45.7-0.4$	&	...			&	22				&	S		&	4.2?		&	...	&	...	\\
G$46.8-0.3$	&	(HC 30)		&	17 $\times$ 13	&	S		&	17		&	...	&	...	\\
G$49.2-0.7$	&	W51C			&	30				&	S? (p)	&	160?		&	Y	&	...	\\
G$54.1+0.3$	&	...			&	12?				&	C?		&	0.5		&	...	&	...	\\
G$54.4-0.3$	&	(HC 40)		&	40				&	S		&	28		&	...	&	Y	\\
G$55.0+0.3$	&	...			&	20 $\times$ 15?	&	S		&	0.5?		&	...	&	...	\\
G$57.2+0.8$	&	(4C 21.53)	&	12?				&	S?		&	1.8		&	...	&	...	\\
G$59.5+0.1$	&	...			&	15				&	S		&	3?		&	...	&	...	\\
G$59.8+1.2$	&	...			&	20 $\times$ 16?	&	?		&	1.5		&	...	&	...	\\
\enddata
\tablenotetext{a}{Sizes taken from Green's SNR Catalog \citep{gre14}.
When it is asymmetric, the major and minor axes of the ellipse are given.}
\tablenotetext{b}{Morphological types of the SNRs in radio observations \citep{gre14}.
The ``S,'' ``F,'' and ``C'' represent ``shell,'' ``filled-center,'' and ``composite'' SNR type, respectively.
The abbreviations within parentheses indicate mixed-morphology (MM) SNRs
that display shell-like morphology in the radio
but filled-center type in X-rays \citep{rho95,rho98,koo16a}:
Y (prototypical MM SNR), p (possible MM SNR).
}
\tablenotetext{c}{Flux density at 1 GHz taken from Green's SNR Catalog \citep{gre14}.}
\tablenotetext{d}{Detection classifications in UWIFE and UWISH2 surveys: Y (detected), ... (not detected).}
\end{deluxetable}

\clearpage
\begin{deluxetable}{lcclc}
\tabletypesize{\scriptsize}
\tablewidth{0pt}
\tablecolumns{5}
\tablecaption{Log of NIR Spectroscopy \label{tab-log}}
\tablehead{
\colhead{Target}   &\colhead{Slit Number}   & \colhead{Slit Position}  &
\colhead{Filter}  & \colhead{Exposure Time}	\\
\colhead{}   &\colhead{}   & \colhead{[ $\alpha$(J2000) ~~~ $\delta$(J2000) ]}  &
\colhead{}  & \colhead{(s)}
}
\startdata
G11.2$-$0.3		&	Slit 1	&	18:11:26.5 $-$19:22:29	&	$K$										&	$120 \times 16$	\\
				&	Slit 2	&	18:11:34.0 $-$19:26:17	&	$K$										&	$120 \times 8$	\\[5pt]
Kes 69			&	Slit 1	&	18:33:12.6 $-$10:12:18	&	$H_{\rm s}$, $K_{\rm s}$					&	$120 \times 2$	\\[5pt]
Kes 73			&	Slit 1	&	18:41:18.3 $-$04:57:02	&	$J_{\rm l}$, $H_{\rm s}$					&	$120 \times 2$	\\[5pt]
3C 391			&	Slit 1	&	18:49:21.7 $-$00:55:34	&	$H_{\rm s}$								&	$120 \times 2$	\\
 				&	Slit 2	&	18:49:33.8 $-$00:56:20	&	$H_{\rm s}$								&	$120 \times 2$	\\
\enddata
\end{deluxetable}

\clearpage
\begin{deluxetable}{cccccccccc}
\tabletypesize{\tiny}
\tablewidth{0pt}
\tablecolumns{10}
\tablecaption{\feii\ and \hh\ Luminosities of 27 Galactic SNRs \label{tab-flux}}
\tablehead{
\colhead{G-Name}   &\colhead{Other Name}  &\colhead{Type} & \colhead{Distance}  &	\colhead{$N_{\rm H}$}  &
\colhead{$F_{\rm [Fe~II]1.644}$\tablenotemark{a}}	&	\colhead{$F_{\rm H_{2}2.122}$\tablenotemark{a}}	&
\colhead{$L_{\rm [Fe~II]1.644}$\tablenotemark{b}} &	\colhead{$L_{\rm H_{2}2.122}$\tablenotemark{b}}	&
\colhead{Reference\tablenotemark{c}}	\\
\cline{6-7}
\cline{8-9}
\colhead{}   & \colhead{}  & \colhead{(MM?)} & \colhead{(kpc)}   & \colhead{($10^{22}$ cm$^{-2}$)}   &
\multicolumn{2}{c}{($10^{-12}~\ergscm$)}	&
\multicolumn{2}{c}{($L_{\odot}$)}
}
\startdata
G$ 8.7-0.1$  & (W30)		 &	S?		& 	 4.5						&	1.2					& 	 9.1		&	 ...		&	17		&	...	&	1		\\
G$ 9.9-0.8$  & ... 		 &	S		& 	 4.0						&	1.3\tablenotemark{e}	&	 ...		& 	 0.48	&	...		&	0.52	&	2		\\
G$11.2-0.3$  & ... 		 &	C		& 	 4.4 					&	3.0					& 	 13		&	 7.4		&	120		&	26	&	3,4,5	\\
G$13.5+0.2$  & ...		 &	S		& 	 13\tablenotemark{d}		&	4.4\tablenotemark{e}	&	 ...		& 	 0.76	&	...		&	54	&	6		\\	
G$15.9+0.2$  & ... 		 &	S?		& 	 10\tablenotemark{d}		&	4.0					& 	 0.082	&	 ...		&	9.4		&	...	&	6,7		\\
G$16.0-0.5$  & ... 		 &	S		& 	 7.5\tablenotemark{d}	&	2.5\tablenotemark{e}	&	 ...		& 	 4.1		&	...		&	32	&	6		\\
G$18.1-0.1$  & ... 		 &	S		& 	 5.6	 					&	1.8					& 	 0.17	&	 1.6		&	0.83		&	4.5	&	8		\\
G$18.9-1.1$  & ... 		 &	C?		& 	 2.0	 					&	1.0					& 	 2.3		&	 2.4		&	0.72		&	0.54	&	9,10		\\
G$21.5-0.9$  & ... 		 &	C		& 	 4.6 					&	2.2					& 	 0.86	&	 ...		&	4.1		&	...	&	3,11		\\
G$21.6-0.8$  & ... 		 &	S		& 	 8.2\tablenotemark{d}	&	2.8\tablenotemark{e}	&	 ...		& 	 0.096	&	...		&	1.0	&	6		\\
G$21.8-0.6$  & Kes 69	 &	S		& 	 5.2						&	2.4					& 	 6.3		&	 13		&	46		&	44	&	12,13	\\
G$23.3-0.3$  & W41		 &	S		& 	 4.2 					&	1.4\tablenotemark{e}	& 	 5.9		&	 ...		&	11		&	...	&	14		\\
G$24.7+0.6$  & ...		 &	C?		& 	 3.5						&	1.2\tablenotemark{e}	&	 ...		& 	 1.8		&	...		&	1.4	&	15		\\
G$27.4+0.0$  & Kes 73	 &	S		& 	 8.5 					&	2.6					& 	 1.8		&	 0.16	&	43		&	1.7	&	16,17	\\
G$27.8+0.6$  & ... 		 &	F		& 	 2.0						&	1.5					& 	 2.0		&	 0.33	&	0.98		&	0.10	&	18,19	\\
G$28.6-0.1$  & ... 		 &	S		& 	 9.6						&	3.5					& 	 2.1		&	 ...		&	140		&	...	&	20,21	\\
G$31.9+0.0$  & 3C 391	 &	S(Y)		& 	 7.1 					&	2.9					& 	 45		&	 11		&	960		&	100	&	22,23	\\
G$32.1-0.9$  & ...		 &	C?		& 	 4.6						&	0.2					&	 ...		& 	 2.8		&	...		&	2.0	&	24		\\
G$32.8-0.1$  & Kes 78 	 &	S?		& 	 4.8						&	0.7					& 	 15		&	 8.6		&	21		&	9.4	&	25,26	\\
G$33.2-0.6$  & ...		 &	S		& 	 7.9\tablenotemark{d}	&	2.7\tablenotemark{e}	&	 ...		& 	 1.4		&	...		&	14	&	6,27		\\
G$34.7-0.4$  & W44		 &	C(Y)		& 	 2.8 					&	1.7					& 	 45		&	 220		&	51		&	150	&	3,28,29	\\
G$39.2-0.3$  & 3C 396 	 &	C(Y)		& 	 8.5						&	4.7					& 	 10		&	 2.7		&	1600		&	97	&	30,31,32	\\
G$41.1-0.3$  & 3C 397 	 &	S(Y)		& 	 10	 					&	3.6					& 	 21		&	 ...		&	1700		&	...	&	33,34,35	\\
G$41.5+0.4$  & ... 		 &	S?		& 	 4.1	 					&	1.4\tablenotemark{e}	& 	 12		&	 ...		&	21		&	...	&	20		\\
G$43.3-0.2$  & W49B		 &	S(Y)		& 	 10 						&	5.0					& 	 53		&	 11		&	15000	&	680	&	36,37,38\\
G$49.2-0.7$  & W51C		 &	S?(p)	& 	 6.0 					&	1.8					& 	 5.6		&	 ...		&	32		&	...	&	39,40,41\\
G$54.4-0.3$  & (HC 40)	 &	S		& 	 6.6						&	2.5					&	 ...		& 	 1.1		&	...		&	6.4	&	22,42	\\
\enddata
\tablenotetext{a}{Detected [Fe II] 1.644~\micron\ and/or \hh\ 2.122~\micron\ fluxes.
The uncertainty given by the quadrature sum of photometric uncertainty
(6\% for [Fe II] and 4\% for \hh) and background RMS noise is less than 10\% of the flux (see the text).}
\tablenotetext{b}{[Fe II] 1.644~\micron\ and/or \hh\ 2.122~\micron\ luminosities after correcting for extinction estimated from $N_{\rm H}$.}
\tablenotetext{c}{References of the adopted distances (fourth column) and column densities (fifth column).}
\tablenotetext{d}{Distance derived from the $\Sigma-D$ relation.}
\tablenotetext{e}{Column density estimated from the mean ratio of visual extinction to path length,
$\left\langle A_{\rm V}/L \right\rangle \approx 1.8$ mag kpc$^{-1}$ \citep{whi92}.}
\tablerefs{
(1) \citet{hew09}, 
(2) \citet{kil16},
(3) \citet{gre04},
(4) \citet{lee13},
(5) \citet{bor16},
(6) \citet{pav14},
(7) \citet{rey06},
(8) \citet{lea14},
(9) \citet{fur89},
(10) \citet{tul10},
(11) \citet{sla00},
(12) \citet{zho09},
(13) \citet{yus03},
(14) \citet{lea08},
(15) \citet{ran18c},
(16) \citet{tia08},
(17) \citet{kum14},
(18) \citet{rei84},
(19) \citet{mis10},
(20) \citet{ran18b},
(21) \citet{uen03},
(22) \citet{ran17},
(23) \citet{sat14},
(24) \citet{fol97},
(25) \citet{zho11},
(26) \citet{bam16},
(27) \citet{par13},
(28) \citet{rho94},
(29) \citet{she04},
(30) \citet{lee09},
(31) \citet{har99},
(32) \citet{su11},
(33) \citet{jia10},
(34) \citet{lea16},
(35) \citet{saf05},
(36) \citet{zhu14},
(37) \citet{keo07},
(38) \citet{hwa00},
(39) \citet{koo95},
(40) \citet{koo05},
(41) \citet{sas14},
(42) \citet{bou05}
}
\end{deluxetable}

\clearpage
\begin{deluxetable}{clccccc}
\tabletypesize{\scriptsize}
\tablewidth{0pt}
\tablecolumns{7}
\tablecaption{\hh\ Emission lines in G11.2$-$0.3 and Kes 69\label{tab-h2prop}}
\tablehead{
\colhead{Wavelength}	& \colhead{Identification}	&	\multicolumn{5}{c}{Relative Flux\tablenotemark{a}} 	\\
\cline{3-7}
\colhead{(\micron)}	& \colhead{ }	&
\colhead{G11.2$-$0.3-N}	& \colhead{G11.2$-$0.3-NE}	&	\colhead{G11.2$-$0.3-SE1}	& \colhead{G11.2$-$0.3-SE2}	&
\colhead{Kes 69-SE}
}
\startdata
1.7480	&	H$_{2}$ 1--0 S(7)	&	...			&	...			&	...			&	...			&	0.23 (0.04)	\\[5pt]
2.0338	&	H$_{2}$ 1--0 S(2)	&	0.27 (0.07)	&	...			&	$< 0.58$\tablenotemark{b}		&	0.49 (0.13)	&	0.28 (0.05)	\\
2.0735	&	H$_{2}$ 2--1 S(3)	&	0.17 (0.03)	&	*			&	*			&	0.23 (0.06)	&	0.16 (0.04)	\\
2.1218	&	H$_{2}$ 1--0 S(1)	&	1.00 (0.03)	&	1.00 (0.05)	&	1.00 (0.06)	&	1.00 (0.03)	&	1.00 (0.03)	\\
2.2233	&	H$_{2}$ 1--0 S(0)	&	0.19 (0.03)	&	0.20 (0.08)	&	$< 0.18$\tablenotemark{b}	&	0.17 (0.05)	&	0.20 (0.03)	\\
2.2477	&	H$_{2}$ 2--1 S(1)	&	0.09 (0.05)	&	$< 0.13$\tablenotemark{b}		&	...			&	...			&	...			\\[5pt]
\hline
\multicolumn{2}{r}{$v_{\rm LSR}$\tablenotemark{c} ($\kms$)} 	&	$+47$ (2)	&	$+45$ (3)	&	$+41$ (3)	&	$+44$ (1)	&	$+57$ (1)	\\
\multicolumn{2}{r}{FWHM\tablenotemark{d} ($\kms$)}			&	$139$ (4)	&	$144$ (6)	&	$139$ (6)	&	$136$ (3)	&	$141$ (3)	\\
\enddata
\tablenotetext{a}{Extinction-corrected fluxes normalized by the H$_{2}$ 1--0 S(1) line
assuming $N_{\rm H}=3.0\times10^{22}$ cm$^{-2}$ for G11.2$-$0.3 \citep{lee13,bor16}
and $N_{\rm H}=2.4\times10^{22}$ cm$^{-2}$ for Kes 69 \citep{yus03},
and the extinction model of the general interstellar dust \citep{dra03}.
The symbol ``...'' indicates that
the lines are located outside of the spectral coverage or detector gap (for \hh\ 2--1 S(1)).
We also mark with an * the lines
that are contaminated by strong OH airglow emission lines
such that we cannot measure their fluxes.
}
\tablenotetext{b}{$3\sigma$ upper limits for the undetected emission line.}
\tablenotetext{c}{Radial velocity of the \hh\ 1--0 S(1) line at the local standard-of-rest frame.
The uncertainty in parentheses is the $1\sigma$ statistical error by a single Gaussian fitting
and does not include the wavelength-calibration error, which is roughly $3~\kms$.}
\tablenotetext{d}{FWHM of the \hh\ 1--0 S(1) line.
The instrumental profile at $\sim 2.12~\micron$ has an FWHM of $\sim 140~\kms$.}
\end{deluxetable}

\clearpage
\begin{deluxetable}{clcc}
\tabletypesize{\scriptsize}
\tablewidth{0pt}
\tablecolumns{4}
\tablecaption{NIR Emission lines in Kes 73 and 3C 391\label{tab-feprop}}
\tablehead{
\colhead{Wavelength}	& \colhead{Identification}	&	\multicolumn{2}{c}{Relative Flux\tablenotemark{a}} 	\\
\cline{3-4}
\colhead{(\micron)}	& \colhead{(lower--upper)}	&	\colhead{Kes 73-Knot A}	& \colhead{3C 391-Spot A}
}
\startdata
1.1886	&	[P II] ~~ $^{3}P_{2}$~-~$^{1}D_{2}$			&	$<0.13$\tablenotemark{b}	&	...	\\
1.2570	&	[Fe II] ~ $a^{6}D_{9/2}$~-~$a^{4}D_{7/2}$	&	1.42 (0.05)	&	...			\\
1.2707	&	[Fe II] ~ $a^{6}D_{1/2}$~-~$a^{4}D_{1/2}$	&	0.08 (0.04)	&	...			\\
1.2791	&	[Fe II] ~ $a^{6}D_{3/2}$~-~$a^{4}D_{3/2}$	&	0.17 (0.06)	&	...			\\
1.2822	&	H I Pa$\beta$							 	&	0.12 (0.04)	&	...			\\
1.2946	&	[Fe II] ~ $a^{6}D_{5/2}$~-~$a^{4}D_{5/2}$	&	0.14 (0.05)	&	...			\\
1.3209	&	[Fe II] ~ $a^{6}D_{7/2}$~-~$a^{4}D_{7/2}$	&	0.38 (0.06)	&	...			\\[5pt]
1.5339	&	[Fe II] ~ $a^{4}F_{9/2}$~-~$a^{4}D_{5/2}$	&	0.17 (0.02)	&	0.08 (0.02)	\\
1.5999	&	[Fe II] ~ $a^{4}F_{7/2}$~-~$a^{4}D_{3/2}$	&	0.10 (0.01)	&	0.04 (0.02)	\\
1.6440	&	[Fe II] ~ $a^{4}F_{9/2}$~-~$a^{4}D_{7/2}$	&	1.00 (0.02)	&	1.00 (0.02)	\\
1.6642	&	[Fe II] ~ $a^{4}F_{5/2}$~-~$a^{4}D_{1/2}$	&	0.05 (0.01)	&	0.02 (0.01)	\\
1.6773	&	[Fe II] ~ $a^{4}F_{7/2}$~-~$a^{4}D_{5/2}$	&	0.11 (0.01)	&	0.06 (0.01)	\\
1.7116	&	[Fe II] ~ $a^{4}F_{5/2}$~-~$a^{4}D_{3/2}$	&	0.02 (0.01)	&	$<0.03$\tablenotemark{b}	\\
1.7976	&	[Fe II] ~ $a^{4}F_{3/2}$~-~$a^{4}D_{3/2}$	&	0.04 (0.01)	&	0.03 (0.01)	\\
1.8099	&	[Fe II] ~ $a^{4}F_{7/2}$~-~$a^{4}D_{7/2}$	&	0.26 (0.05)	&	0.25 (0.04)	\\[5pt]
\hline
\multicolumn{2}{r}{$v_{\rm LSR}$\tablenotemark{c} ($\kms$)} 	&	$-30$ (1)	&	$+11$ (1)\\
\multicolumn{2}{r}{FWHM\tablenotemark{d} ($\kms$)}			&	$178$ (2)	&	$162$ (2)\\
\enddata
\tablenotetext{a}{Extinction-corrected fluxes normalized by the [Fe II] 1.644~\micron\ line
assuming $N_{\rm H}=2.6\times10^{22}$ cm$^{-2}$ for Kes 73 \citep{kum14}
and $N_{\rm H}=2.9\times10^{22}$ cm$^{-2}$ for 3C 391 \citep{sat14},
and the extinction model of the general interstellar dust \citep{dra03}.
The symbol ``...'' indicates that
the lines are located outside of the spectral coverage.}
\tablenotetext{b}{$3\sigma$ upper limits for the undetected emission line.}
\tablenotetext{c}{Radial velocity of the [Fe II] 1.644~\micron\ line at the local standard-of-rest frame.
The uncertainty in parentheses is the $1\sigma$ statistical error by a single Gaussian fitting
and does not include the wavelength-calibration error, which is roughly $3~\kms$.}
\tablenotetext{d}{FWHM of the [Fe II] 1.644~\micron\ line.
The instrumental profile at $\sim 1.64~\micron$ has an FWHM of $\sim 130~\kms$.}
\end{deluxetable}


\clearpage
\begin{figure}
\center{
\includegraphics[width=0.9\textwidth]{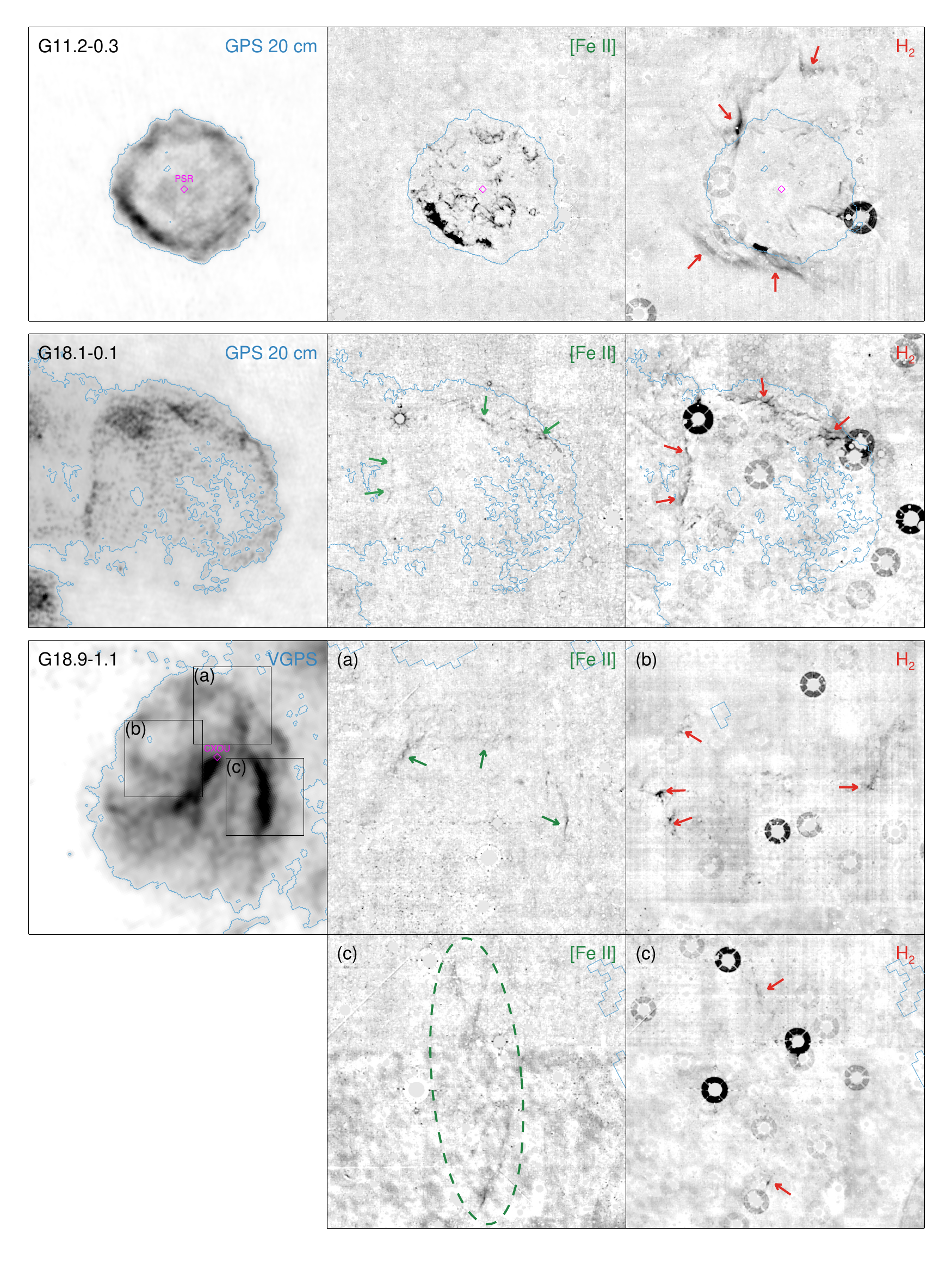}
}
\caption{
   Continuum-subtracted narrowband images of 11 Galactic SNRs
   emitting both \feii\ and \hh\ emission lines.
   The radio images were taken from either
   the VLA 20/90~cm Galactic Plane Survey \citep[GPS 20/90cm;][]{hel06} or
   the VLA 21~cm Galactic Plane Survey \citep[VGPS;][]{sti06}.
   The magenta symbols represent
   the locations of X-ray/radio sources associated with the SNRs:
   crosses = OH masers;
   open diamonds = pulsars or point-like X-ray sources.
} \label{fig-img1-1}
\end{figure}

\addtocounter{figure}{-1}
\clearpage
\begin{figure}
\center{
\includegraphics[width=0.9\textwidth]{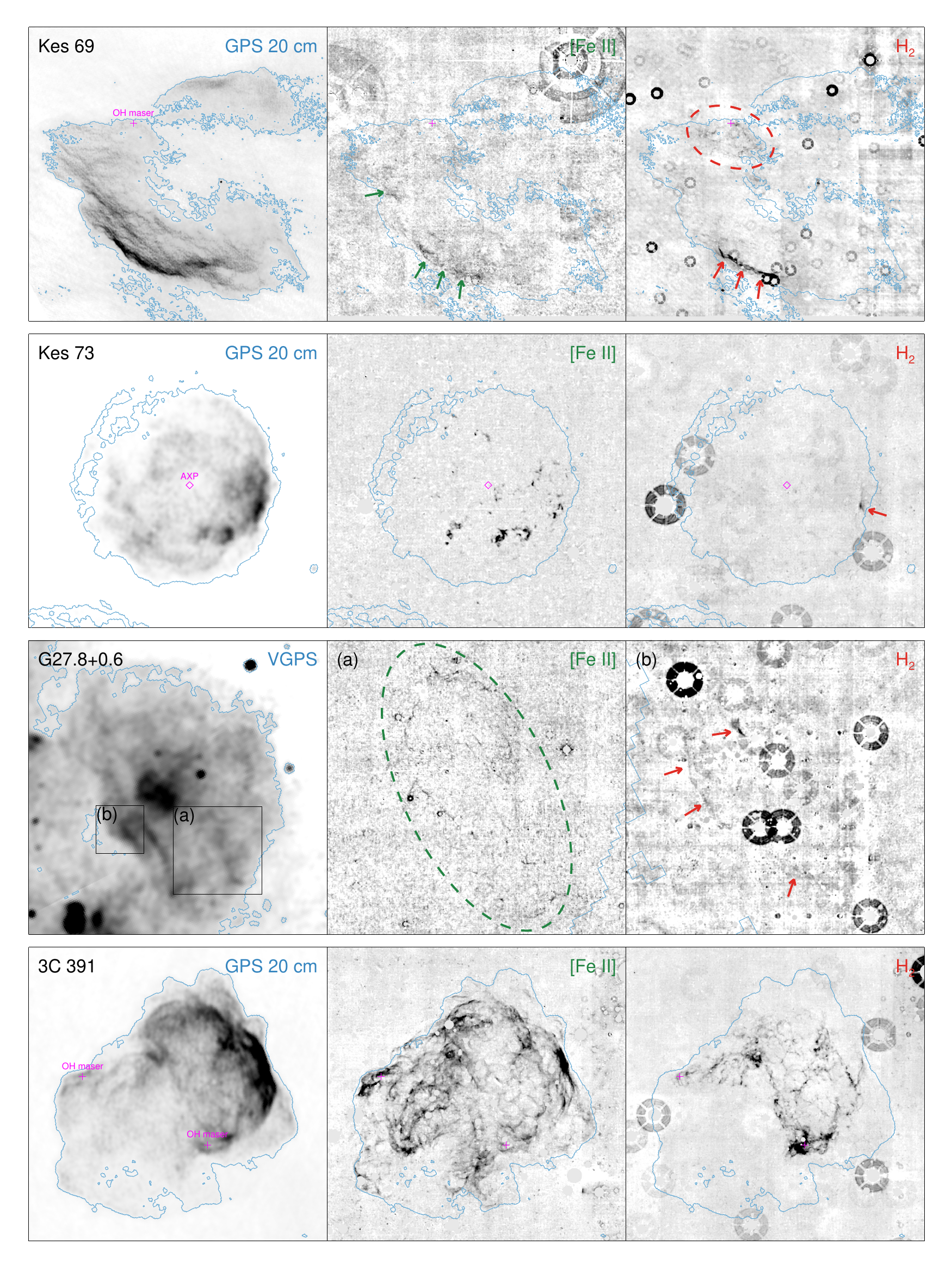}
}
\caption{
   (Continued)
} \label{fig-img1-2}
\end{figure}

\addtocounter{figure}{-1}
\clearpage
\begin{figure}
\center{
\includegraphics[width=0.9\textwidth]{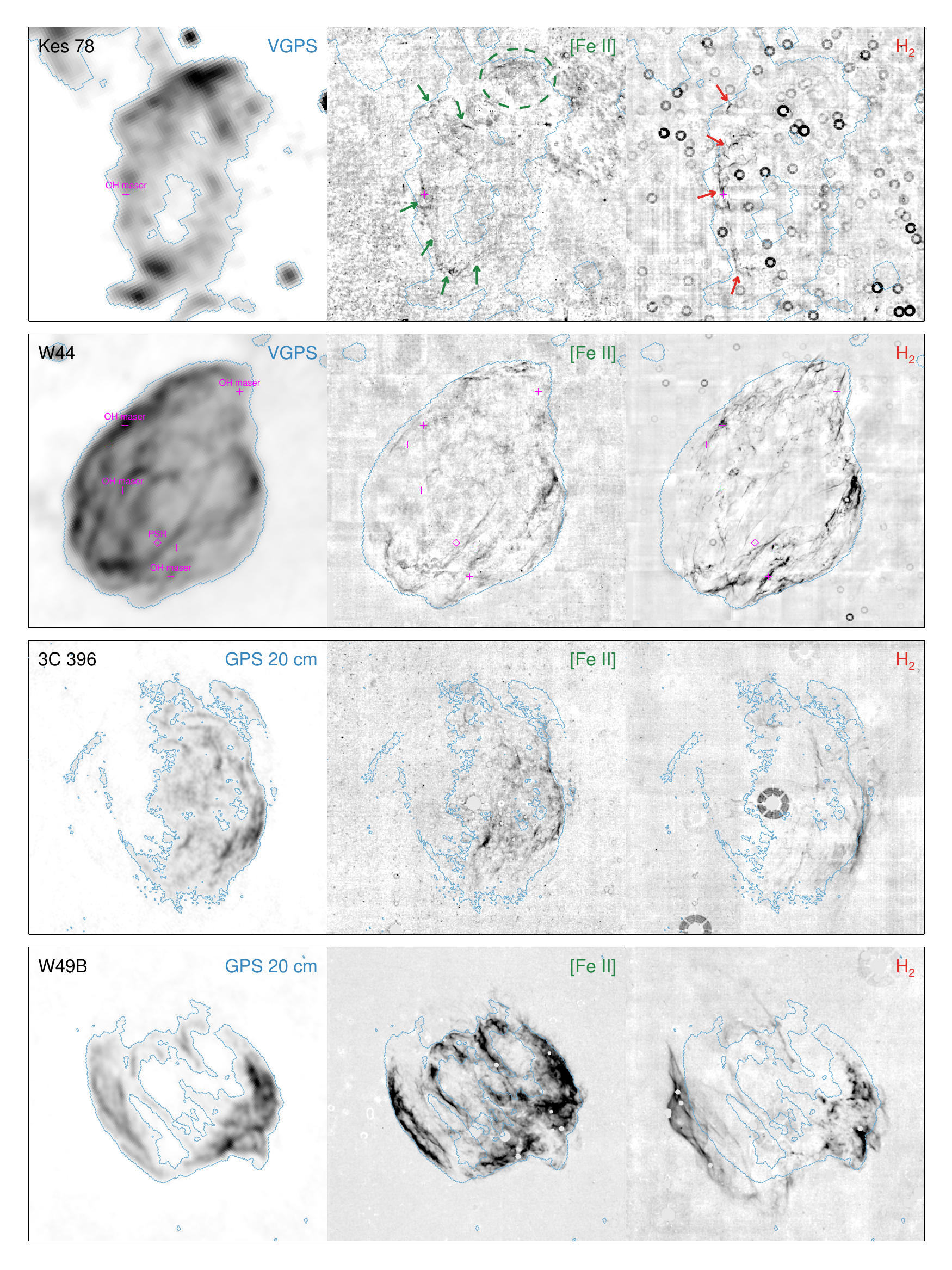}
}
\caption{
   (Continued)
} \label{fig-img1-3}
\end{figure}

\clearpage
\begin{figure}
\center{
\includegraphics[width=\textwidth]{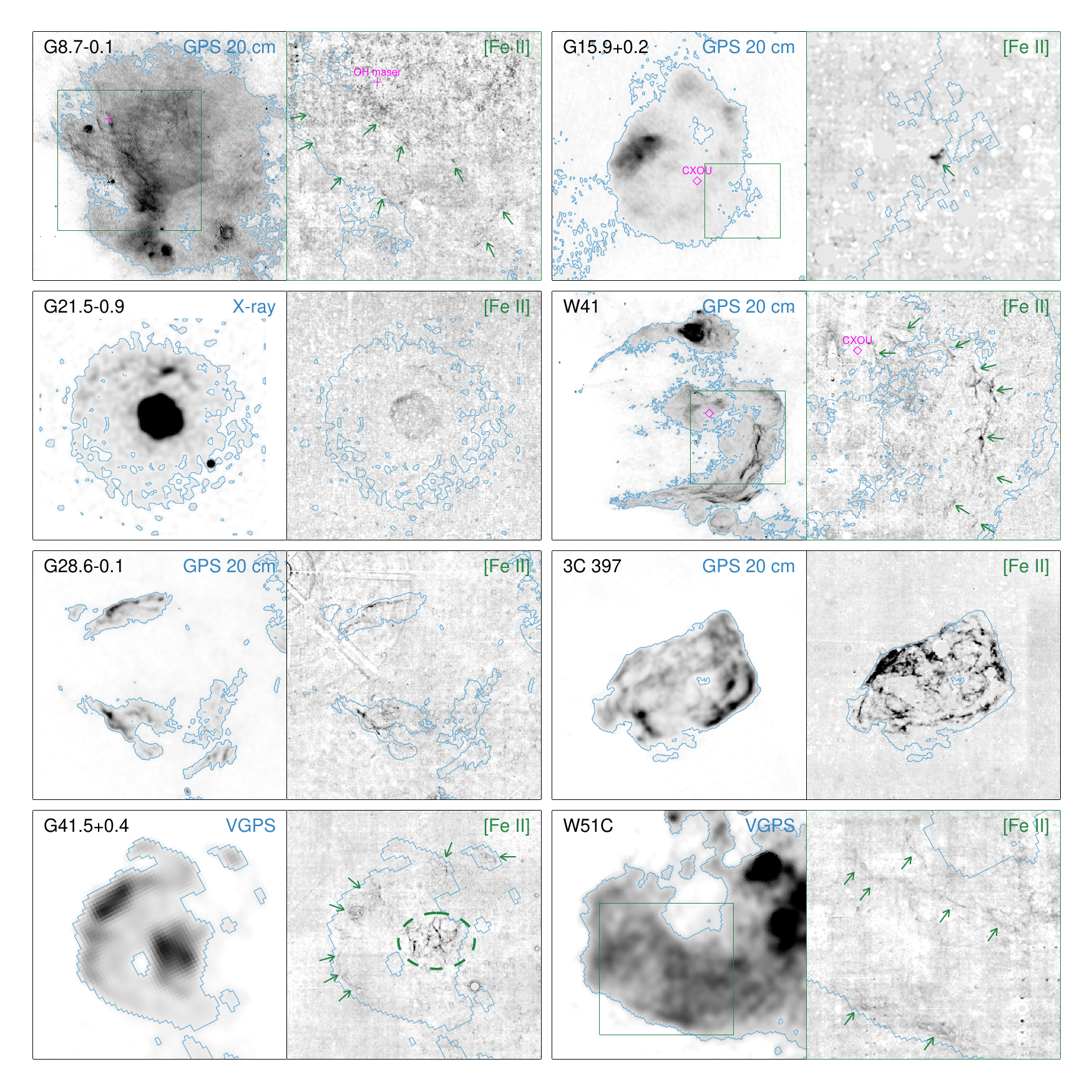}
}
\caption{
   Continuum-subtracted narrowband images of eight Galactic SNRs
   emitting \feii\ emission lines only.
   The radio images were taken from either
   the VLA 20/90~cm Galactic Plane Survey \citep[GPS 20/90cm;][]{hel06} or
   the VLA 21~cm Galactic Plane Survey \citep[VGPS;][]{sti06},
   whereas the X-ray image of G21.5-0.9 was taken from
   the previous {\it Chandra} X-ray observation \citep{sla00}.
   The magenta symbols represent
   the locations of X-ray/radio sources associated with the SNRs:
   crosses = OH masers;
   open diamonds = pulsars or point-like X-ray sources.
} \label{fig-img2}
\end{figure}

\clearpage
\begin{figure}
\center{
\includegraphics[width=\textwidth]{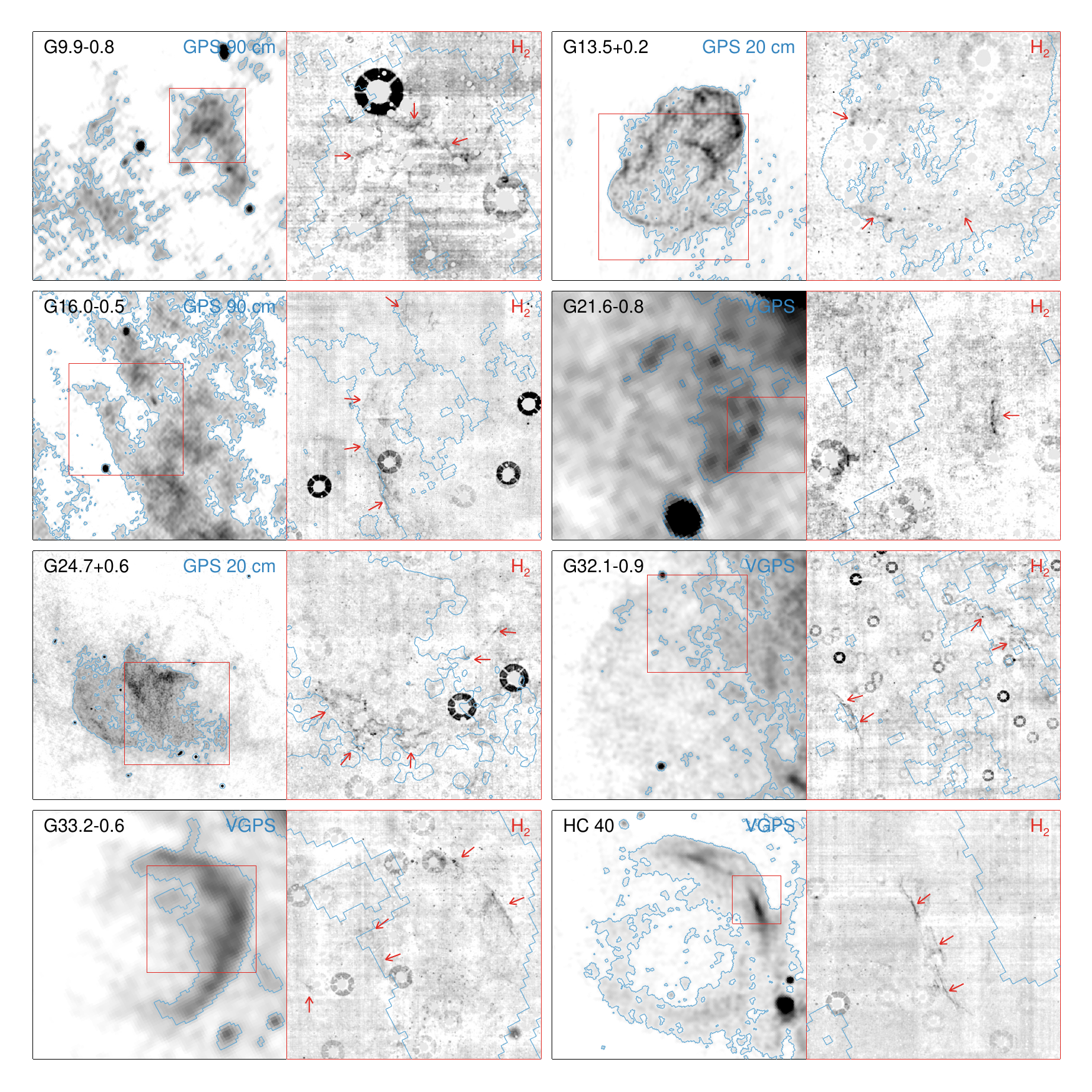}
}
\caption{
   Continuum-subtracted narrowband images of eight Galactic SNRs
   emitting \hh\ emission lines only.
   The radio images were taken from either
   the VLA 20/90~cm Galactic Plane Survey \citep[GPS 20/90cm;][]{hel06} or
   the VLA 21~cm Galactic Plane Survey \citep[VGPS;][]{sti06}.
} \label{fig-img3}
\end{figure}

\clearpage
\begin{figure}
\center{
\includegraphics[width=\textwidth]{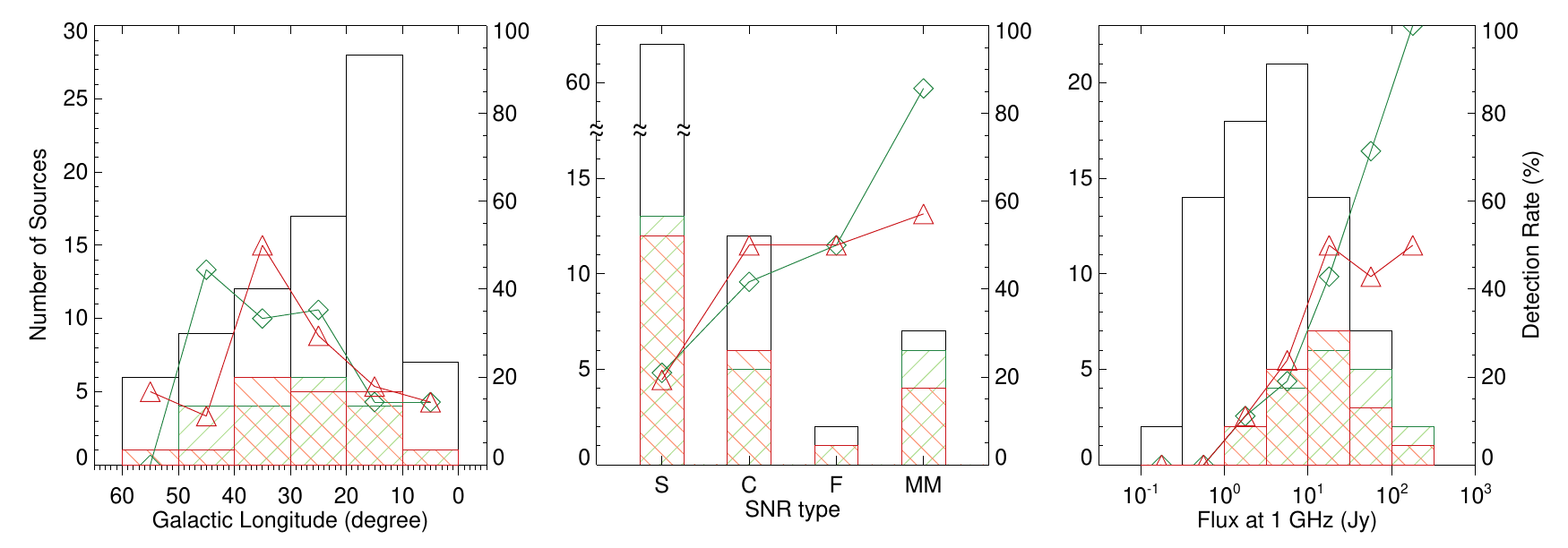}
}
\caption{
   Detection rates vs.
   (left) Galactic longitude, (middle) SNR type,
   and (right) flux density at 1 GHz.
   The black histogram represents the total number of SNRs,
   whereas the green and red hatched histograms denote the number of
   \feii-emitting and \hh-emitting SNRs, respectively.
   The green diamonds and red triangles are the detection rate at each bin.
   Morphological types, ``S,'' ``C,'' ``F'' in the middle panel indicate
   Shell-, Composite-, and Filled-centered SNRs defined by their radio continuum,
   whereas ``MM'' is Mixed-Morphology or thermal-composite type
   showing a shell-like morphology in radio with filled-centered X-ray.
} \label{fig-hist1}
\end{figure}

\clearpage
\begin{figure}
\center{
\includegraphics[width=\textwidth]{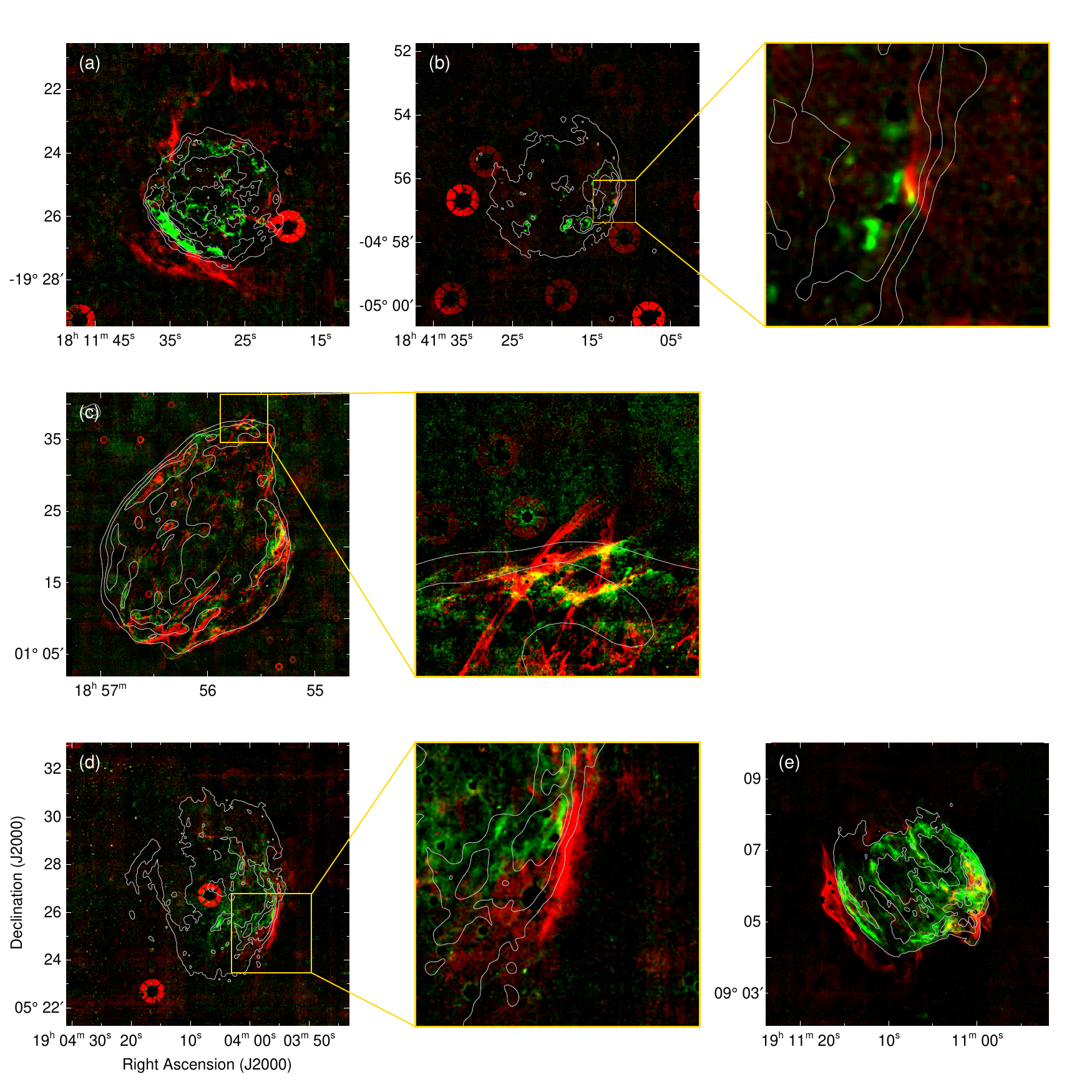}
}
\caption{
   Color-composite images of five Galactic SNRs showing the ``\feii--\hh\ reversal'':
   (a) G11.2-0.3, (b) Kes 73, (c) W44, (d) 3C 396, and (e) W49B.
   The continuum-subtracted \feii\ and \hh\ images are displayed in
   green and red, respectively,
   and the gray contours represent the radio continuum.
} \label{fig-rev}
\end{figure}

\clearpage
\begin{figure}
\center{
\includegraphics[width=0.5\textwidth]{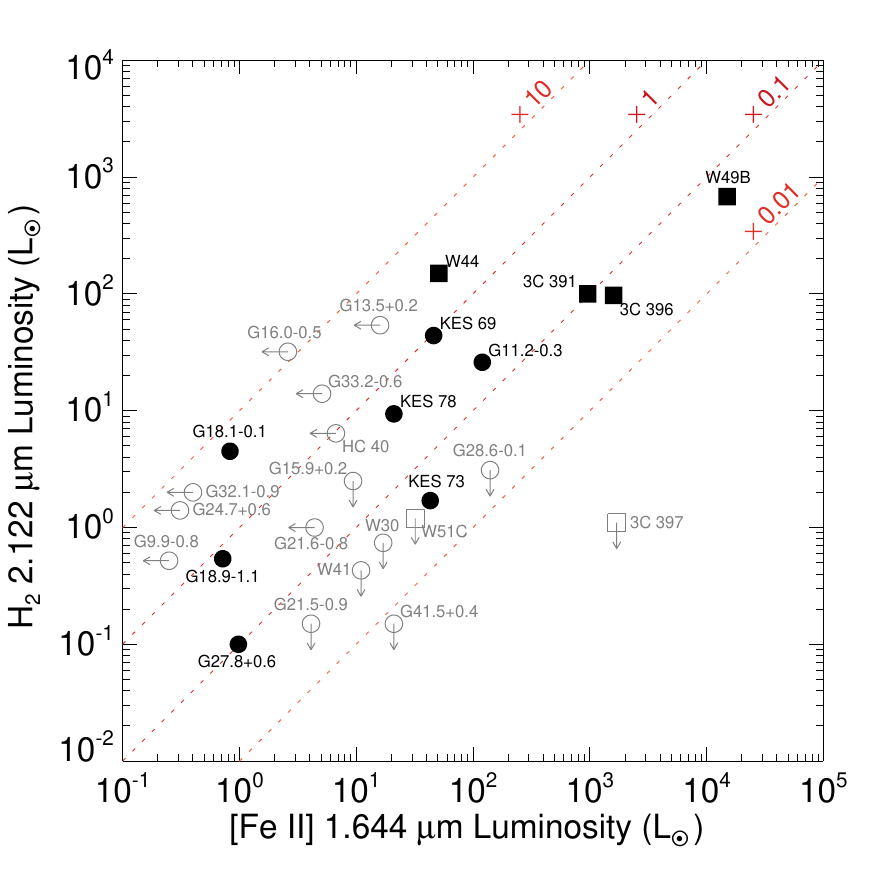}
}
\caption{
   \feii\ 1.644~\micron\ vs. \hh\ 2.122~\micron\ luminosity for 27 SNRs
   emitting \feii/\hh\ emission lines.
   The arrows indicate $3\sigma$ upper limits of the luminosities
   when the emission line is undetected.
   The mixed-morphology SNRs are marked with square symbols.
   The red dashed lines indicate
   the luminosity ratio of the \feii\ and \hh\ emission lines.
} \label{fig-rat}
\end{figure}

\clearpage
\begin{figure}
\center{
\includegraphics[width=\textwidth]{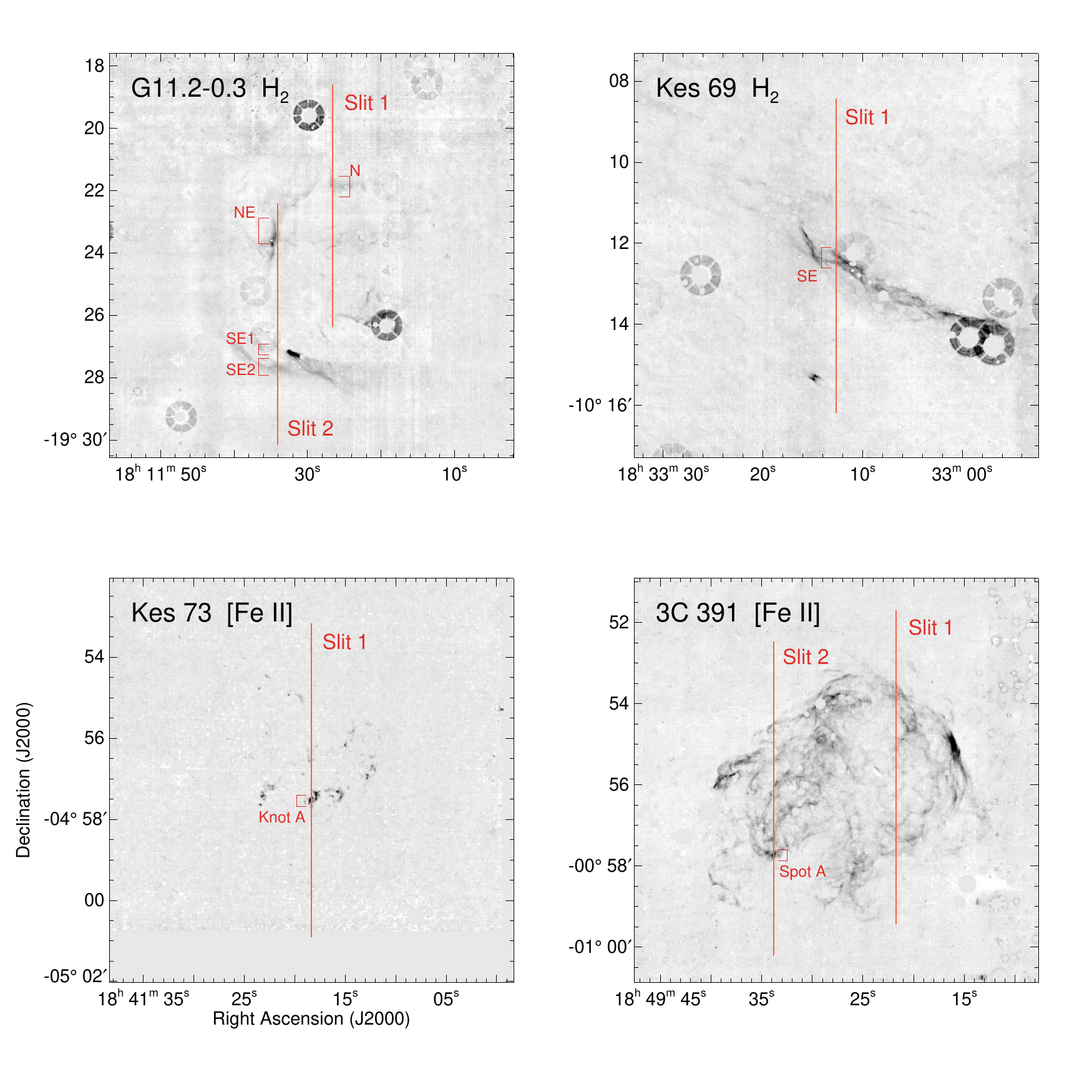}
}
\caption{
   Slit positions (red vertical bars) of the NIR spectroscopy for four SNRs.
   The background images are the continuum-subtracted \hh\ of G11.2$-$0.3 and Kes 69,
   and \feii\ of Kes 73 and 3C 391.
   The source names and their positions are also marked.
} \label{fig-slit}
\end{figure}

\clearpage
\begin{figure}
\center{
\includegraphics[width=\textwidth]{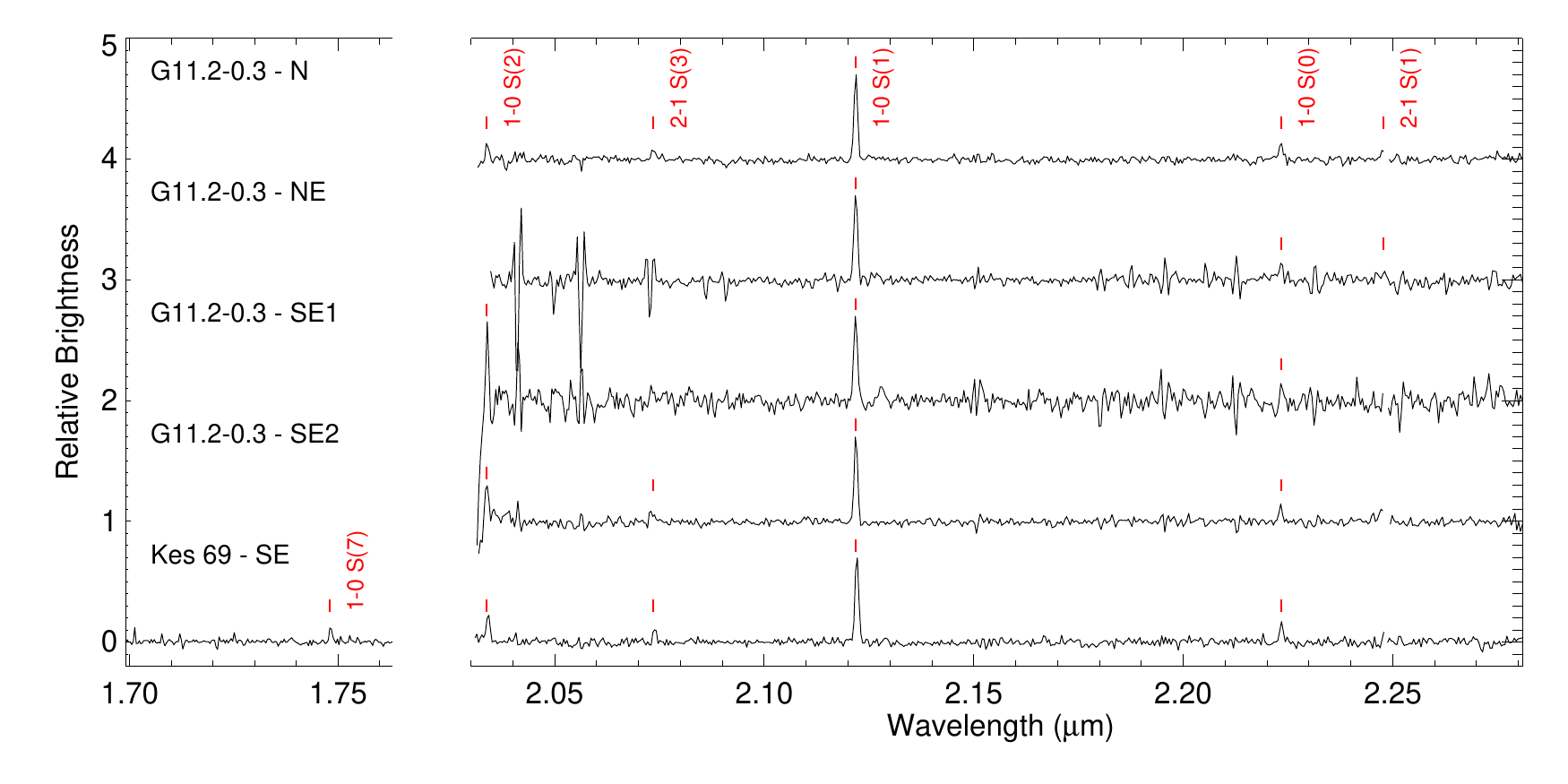}
}
\caption[]{
   NIR $H$- and $K$-band spectra of \hh\ filaments detected in G11.2$-$0.3 and Kes 69.
} \label{fig-specks}
\end{figure}

\clearpage
\begin{figure}
\center{
\includegraphics[width=\textwidth]{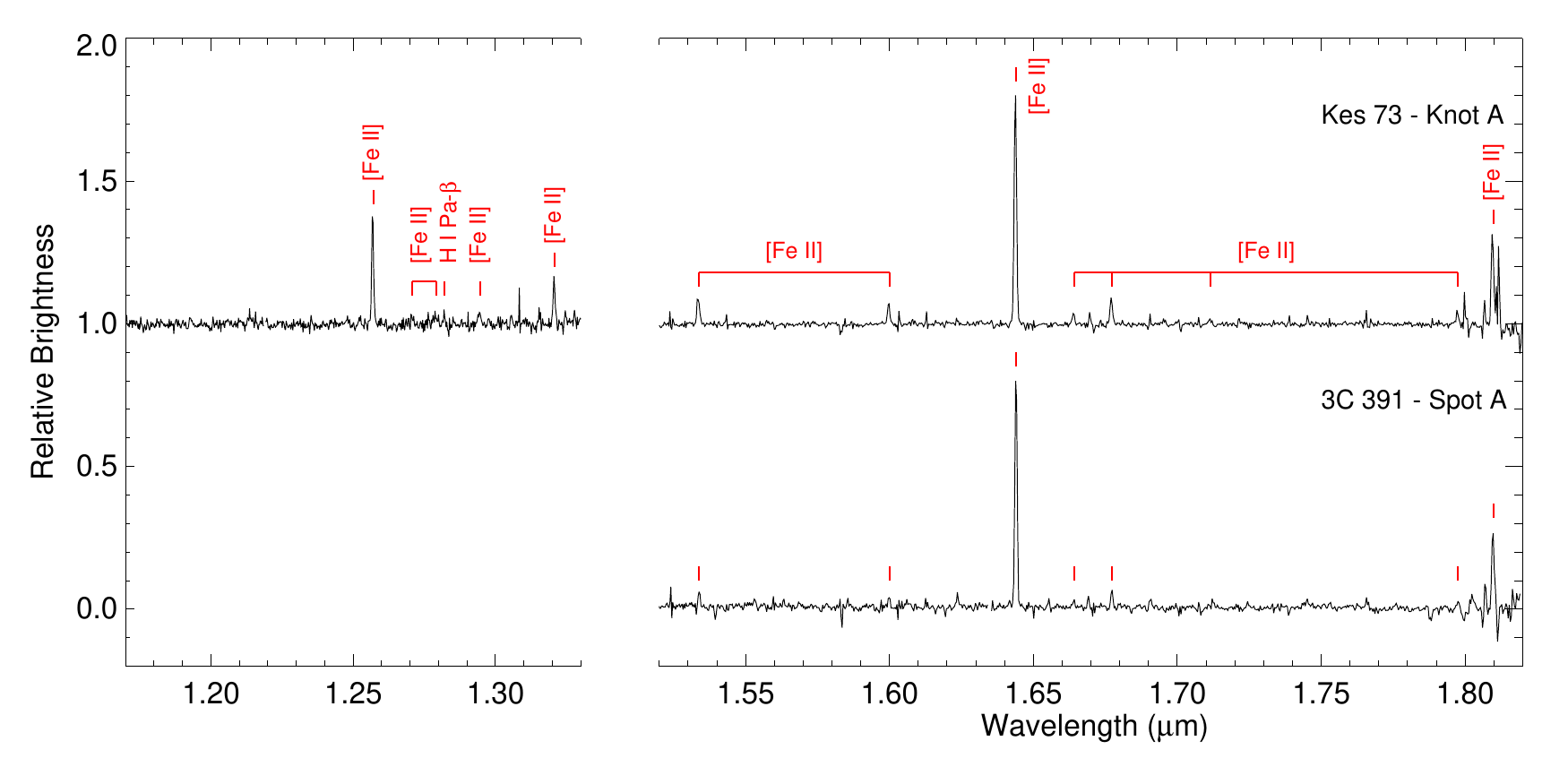}
}
\caption[]{
   NIR $J$- and $H$-band spectra of \feii\ clumps detected in Kes 73 and 3C 391.
} \label{fig-specjh}
\end{figure}

\clearpage
\begin{figure}
\center{
\includegraphics[width=0.5\textwidth]{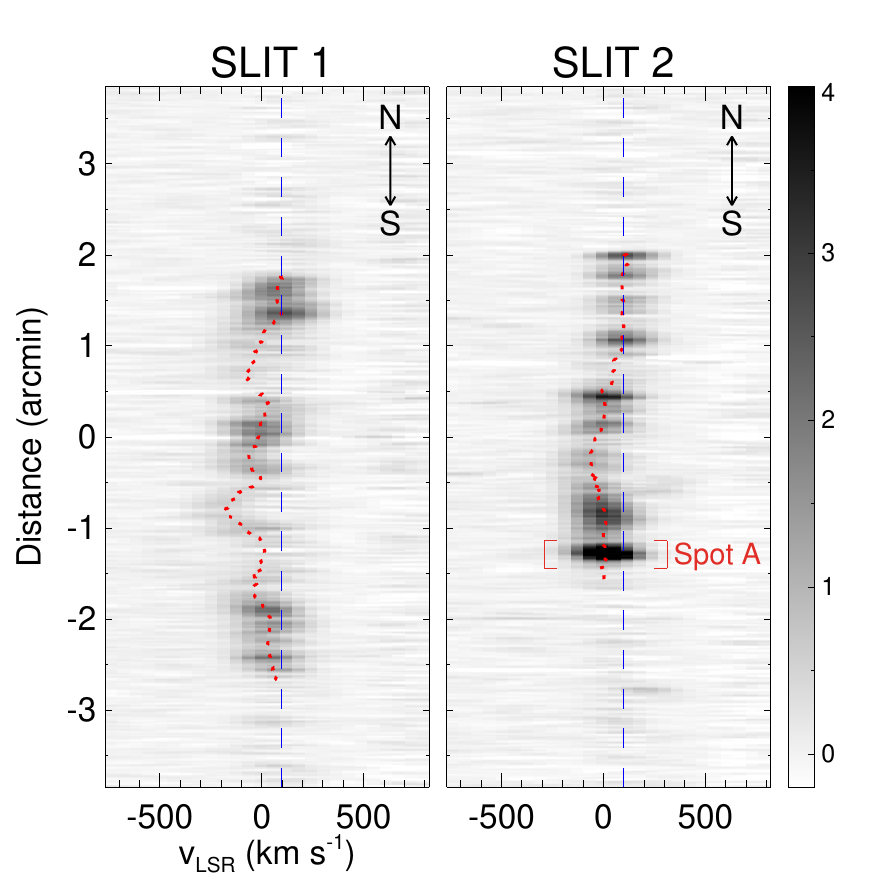}
}
\caption[]{
   Position--velocity diagrams of the \feii\ 1.644~\micron\ lines
   detected in 3C 391.
   The red dotted lines indicate the central velocity of the lines,
   whereas the blue dashed line represents the systematic velocity of the remnant,
   $v_{\rm LSR}=+100~\kms$.
   The slit direction is marked in the upper right of each panel.
   The units of the color bar is 10$^{-17}$ erg s$^{-1}$ cm$^{-2}$ \AA$^{-1}$.
   The 1D spectrum of Spot A in Slit 2 is shown in Figure~\ref{fig-specjh}.
} \label{fig-3c391}
\end{figure}

\clearpage
\begin{figure}
\center{
\includegraphics[width=0.5\textwidth]{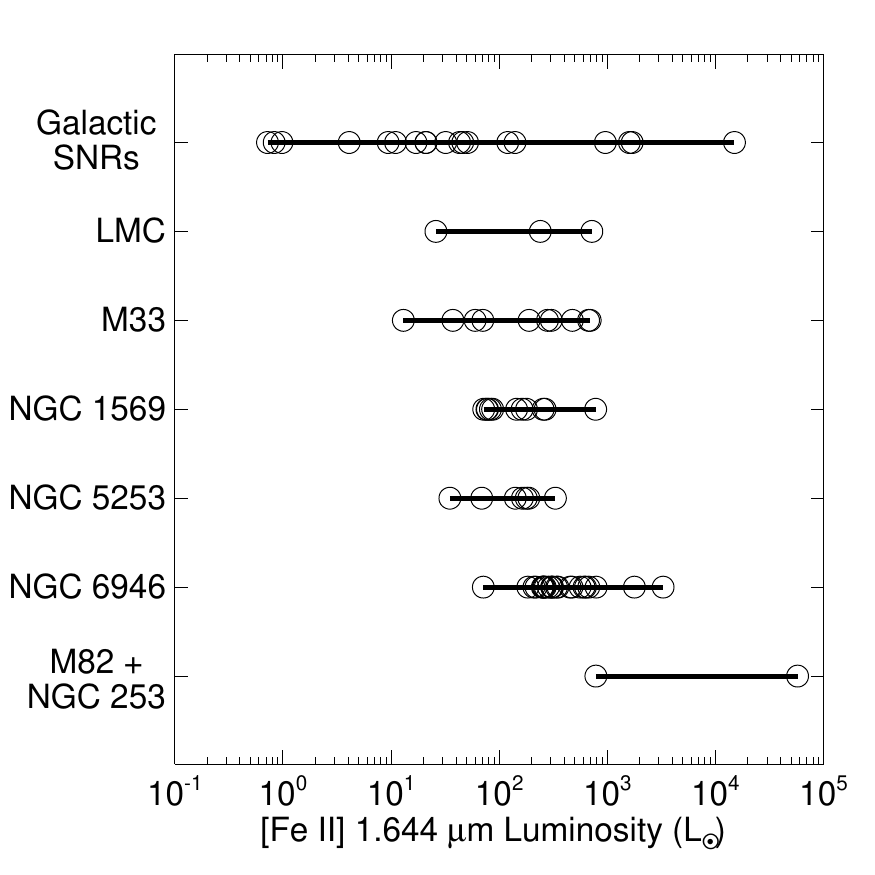}
}
\caption[]{
   \feii\ 1.644~\micron\ luminosity of Galactic and extragalactic SNRs:
   Galactic SNRs (this paper), LMC SNRs \citep{oli89},
   M33 SNRs \citep{mor02}, NGC 1569 and NGC 5253 SNRs \citep{lab06},
   NGC 6946 SNRs \citep{bru14}, and M82 and NGC 253 SNRs \citep{alo03}.
   For M82 and NGC 253, the upper and lower luminosities are presented.
} \label{fig-feii}
\end{figure}

\end{document}